\documentclass[aps,prb,twocolumn]{revtex4}
\usepackage{graphicx}
\usepackage{amsmath}
\usepackage{tikz}
\PassOptionsToPackage{caption=false}{subfig}
\usepackage{subfig}

\makeatletter
\renewcommand{\maketag@@@}[1]{\hbox{\m@th\normalsize\normalfont#1}}%
\makeatother

\newcommand{\bra}[1]{\ensuremath{\langle{#1}|\,}}
\newcommand{\ket}[1]{\ensuremath{\,|{#1}\rangle}}
\newcommand{\braket}[1]{\ensuremath{\langle{#1}\rangle}}

\begin{document}
\title{Distribution of waiting times between electron cotunnelings}
\author{Samuel L. Rudge}
\author{Daniel S. Kosov}
\address{College of Science and Engineering, James Cook University, Townsville, QLD, 4811, Australia }

\begin{abstract}
In the resonant tunneling regime sequential processes dominate single electron transport through quantum dots or molecules that are weakly coupled to macroscopic electrodes. In the Coulomb blockade regime, however, cotunneling processes dominate. Cotunneling is an inherently quantum phenomenon and thus gives rise to interesting observations, such as an increase in the current shot noise. Since cotunneling processes are inherently fast compared to the sequential processes, it is of interest to examine the short time behaviour of systems where cotunneling plays a role, and whether these systems display nonrenewal statistics. We consider three questions in this paper. Given that an electron has tunneled from the source to the drain via a cotunneling  or sequential process, what is the waiting time until another electron cotunnels from the source to the drain? What are the statistical properties of these waiting time intervals? How does cotunneling affect the statistical properties of a system with strong inelastic electron-electron interactions? In answering these questions, we extend the existing formalism for waiting time distributions in single electron transport to include cotunneling processes via an $n$-resolved Markovian master equation. We demonstrate that for a single resonant level the analytic waiting time distribution including cotunneling processes yields information on individual tunneling amplitudes. For both a SRL and an Anderson impurity deep in the Coulomb blockade there is a nonzero probability for two electrons to cotunnel to the drain with zero waiting time inbetween. Furthermore, we show that at high voltages cotunneling processes slightly modify the nonrenewal behaviour of an Anderson impurity with a strong inelastic electron-electron interaction.
\end{abstract}

\maketitle
 
\section{Introduction}

With the ever-present search for smaller transistors and the advent of modern technologies such as quantum computing, the world in recent years has turned its gaze inwards to probe electron transport through nanoscale devices, where a fundamental understanding of quantum dynamics is required. This has yielded intriguing experimental and theoretical results: for example, single-molecule transistors, quantum heat engines, and spintronics.\cite{Tan2017,Bogani2008,Xiang2016} Of particular interest in quantum nanoscale systems is the potential for encountering microscopic current fluctuations and phenomena that are classically forbidden, such as the existence of electron transport through virtual quantum states that temporarily violate energy conservation laws; both of which form the focus of this paper.

Electron transport through quantum systems can display a phenomena known as $\text{\it{cotunneling}}$.  Inelastic cotunneling was first proposed theoretically by Averin and Odintsov \cite{Averin1989} and confirmed experimentally shortly after the theoretical prediction by Geerligs et al.,\cite{PhysRevLett.65.3037} with the 
introduction of the modern combined inelastic and elastic theory detailed simultaneously by Averin and Nazarov.\cite{Averin1990} In contrast to $\text{\it{sequential tunneling}}$, which describes single electron tunneling events and can essentially be described classically, cotunneling is a coherent quantum process that involves the tunneling of an electron from the source to the drain (or vice versa) through an intermediate ``virtual'' state, which may or may not be classically forbidden.\cite{scheer2010molecular,Bruus2002,Beenakker1991} Elastic cotunneling leaves the system with the same energy, while inelastic cotunneling leaves the intermediate quantum system in an excited state. The common explanation is that cotunneling is an example of the uncertainty principle $\Delta t\Delta E \sim\hbar$; energy conservation can be violated only if the electron spends a sufficiently short time in the intermediate virtual state. Although in recent years this notion has been challenged by Romito and Gefen.\cite{Romito2014}

Cotunneling processes dominate transport in the Coulomb blockade regime, as the electronic energy levels are pushed outside the voltage bias window and sequential tunneling is exponentially suppressed. Hence, cotunneling manifests experimentally as a  small current in the Coulomb blockade regime, and as a small correction to the sequential current in the resonant tunneling regime.\cite{PhysRevLett.86.878,PhysRevLett.65.3037} Theoretical research into cotunneling has investigated its effect on transport in systems with inelastic scatterings, such as  electron-electron and electron-phonon interactions.\cite{Koch2006,Koch2004,Turek2002} Additional cotunneling research has focused on heat conductance,\cite{Dinaii2014,Gergs2015} transport in double quantum dots,\cite{Golovach2004,Pedersen2007} and inelastic cotunneling spectroscopy.\cite{Begemann2010,Leijnse2009} Recently multiple authors have studied the noise and full counting statistics (FCS) of cotunneling phenomena in an attempt to explore its effect on current fluctuations.\cite{Leijnse2008,PhysRevB.91.235413,PhysRevB.80.235306,PhysRevLett.95.146806,Sukhorukov2000,Thielmann2005,Weymann2008,Carmi2012,Aghassi2008,Braggio2006,Utsumi2006} Such investigations have demonstrated that inelastic cotunneling transport induces super-Poissonian shot noise for a variety of systems, which is in agreeance with experimental measurements.\cite{OKazaki2013,Zhang2007,ONac2005} 

Alongside the zero-frequency noise and FCS, the waiting time distribution (WTD) has been shown to be a useful tool for describing current fluctuations in quantum nanoscale systems, as it contains information complementary to that found in other statistics.\cite{Brandes2008,PhysRevA.39.1200,Srinivas2010,Thomas2013,Rudge2016,Rudge2016a,Kosov2016,Dasenbrook2015,Albert2011,Albert2012} In contrast to current cumulants, which require theoretical calculations over long time intervals, WTDs can reveal interesting short time physics that may otherwise be inaccessible. Of particular interest is observing a violation of renewal statistics, where the assumption is that $w(\tau_{1},\tau_{2})=w(\tau_{1})w(\tau_{2})$. Nonrenewal statistics is characterised by short time correlations between subsequent waiting times, and is thus invisible in the current cumulants. Perhaps the recent interest has been spurred onwards in part by the development of real-time single electron detection techniques, which have enabled experimental measurement of microscopic current fluctuations for many different quantum systems.\cite{Ubbelohde2012,Lu2003,Gustavsson2009,Gustavsson2008a} However, there remains experimental difficulties in measuring electron tunnelings via virtual processes due to the collapse of the intermediate state.\cite{PhysRevB.78.155309,Zilberberg2014} A possible method for experimentally accessing waiting times including quantum processes is the reconstruction of the WTD from low order charge correlation functions.\cite{Haack2015} Although there are multiple definitions of the WTD in statistics,\cite{VanKampen1981} in the context of quantum transport it is the conditional probability density that, given an extra electron was counted in the drain electrode at time $t$, another extra electron was counted in the drain at time $t+\tau$, where no intermediate tunneling events to the drain are allowed. 

Historically, WTDs have been extensively used in quantum optics as a statistical tool\cite{PhysRevA.39.1200,Srinivas2010} and they were introduced to mesoscopic quantum transport by Brandes, who calculated WTDs by defining jump operators from a quantum master equation.\cite{Brandes2008} The master equation method for calculating WTDs has since been applied to a diverse range of scenarios, such as systems with electron-electron interactions, electron-phonon interactions, coherent internal transport, non-Markovian quantum transport and spintronics.\cite{Thomas2013,Rudge2016,Rudge2016a,Kosov2016,Brandes2008,Rajabi2013,Chevallier2016,Walldorf2018,Sothmann2014,Welack2009,Ptaszynski2017} Alongside the master equation approach, there exists various techniques for calculating WTDs in mesocopic transport. For example, Albert et al.\cite{Albert2012} developed a scattering matrix approach suitable for fully coherent transport, and described single channel and multichannel transport,\cite{Albert2012,Haack2014} transport through superconducting junctions,\cite{Chevallier2016,Albert2016} and transport of electron pulses.\cite{Albert2014,Albert2011} Despite this success, the scattering matrix approach is unable to calculate WTDs outside of the steady state; thus, nonequilibrium Green's functions are used to describe coherent transport in the transient regime.\cite{Tang2014,Tang2014a,PhysRevB.92.125435}  However, so far WTDs have not been used to study the statistical properties of electron cotunneling events for systems with strong electron-electron interactions. Although the scattering matrix approach and nonequilibrium Green's functions are applicable to coherent transport, and thus seem tailor-made for describing cotunneling, they struggle to include strong inelastic scatterings in the quantum system; thus, in this paper we use the master equation technique.

We study the WTD in an Anderson impurity for successive tunnelings to the drain, including cotunneling, and compare it to the WTD for successive tunnelings to the drain for only sequential tunneling processes. We first develop a systematic method for extending the current master equation approach for WTDs developed by Brandes\cite{Brandes2008} to include cotunneling processes, and then demonstrate its use for transport through an Anderson impurity, as well as the limiting case of strong Coulomb repulsion and no level splitting when the system behaves as a single resonant level (SRL). In doing so, we examine the relationship between inelastic scatterings and the inherently coherent quantum cotunneling process, as well as the effect cotunneling has on nonrenewal statistics.

The master equation approach to quantum transport is a powerful method for analysing quantum electron transport through mesoscopic systems. \cite{Gurvitz1996,PhysRevB.71.205304,PhysRevB.74.235309,dzhioev11a,kosov18} Although the full master equation is useful for describing quantum effects such as interference,\cite{Nazarov1993} decoherence between double quantum dots,\cite{Tu2008} electron  transport through quantum dot attached to superconducting leads,\cite{kosov13,PhysRevB.87.155439}
and driven quantum transport,\cite{Kohler2005} in many cases the transport is incoherent and thus is effectively described by rate equations;\cite{Gurvitz1998,Gurvitz1996a,Davies1993} this is the approach taken in this paper. In order to connect this formalism to waiting times we will in fact have to work with the $\text{\it{n}}$-resolved master equation.\cite{Li2005,Emary2007,Gurvitz1996a}

The transition rates in the master equation are calculated using the T-matrix approach: a perturbation expansion around the tunneling coupling $H_T$. Sequential tunneling corresponds to the lowest order of this expansion, and cotunneling processes correspond to next-to-leading order in $H_{T}$: first and second order in the tunneling coupling strength $\gamma$, respectively. Cotunneling rates developed from a purely second-order perturbative expansion about $H_{T}$ are well-known to formally diverge due to higher-order tunneling effects not being taken into account. To overcome this we follow the approach first developed by Averin\cite{Averin1994}, and extended to the T-matrix context by Turek and Matveev\cite{Turek2002} and Koch et al.,\cite{Koch2004,Koch2006} of introducing a finite width to the energy of the  intermediate virtual state. Specifically, we closely follow the methodology of Koch et al. and obtain similar analytic results, although we note that we focus on the WTDs associated with an electron-electron interaction whereas Koch et al. focus on an electron-vibration interaction in the limit $U\rightarrow\infty$. Once the rates are defined, and similarly to Thomas and Flindt's approach, we start with an $\text{\it{n}}$-resolved master equation, then derive the WTD from the idle time probability and show that for forward tunneling only it reduces to the method introduced by Brandes, albeit with a non-intuitive Liouvillian splitting.\cite{Thomas2013}

We demonstrate that, likewise to the WTD for sequential tunneling through a single resonant level, the WTD including cotunneling offers information on the individual electrode coupling parameters.\cite{Brandes2008} Furthermore, for an Anderson impurity cotunneling processes slightly increase the nonrenewal behaviour; this is evident in the comparison of the correlation between subsequent waiting times, which is largely controlled by the strength of the Coulomb repulsion. However, the use of the method presents difficulties in two key areas: when the level is inside the voltage bias window and when backward tunneling processes are included.

The paper is organised as follows. Section \ref{Methods} outlines the construction of the master equation and the derivation of the WTD including cotunneling. Section \ref{Results} details analytic results for cotunneling through an Anderson impurity and a SRL. Section \ref{Conclusion} outlines the main results and discusses future work. The Appendix details calculations and derivations used throughout the paper.

Throughout this paper we use natural units: $\hbar=k_e=e=1$.

%%%%%%%%%%%%%%%%%%%%%%%%%%%%%%%%%%%%%%%%%%%%%%%%%%%%%%%%%%%%%%%

\section{Methods} \label{Methods}

\subsection{Quantum rates for cotunneling processes}

In this paper we examine the transport of electrons, modelled as fermions with spin. Let us consider a nanoscale quantum system weakly coupled to two macroscopic metal electrodes: the source and drain. The source and drain are held at different chemical potentials to cause a voltage bias across the system and induce a non-equilibrium state. For such a setup, the Hamiltonian is 

\begin{equation}
H = H_S+H_D+H_M+H_T.
\end{equation}
The source and drain are modelled as a sea of non-interacting electrons with the Hamiltonians 
\begin{equation}
H_S = \sum_{s,\sigma}\varepsilon_{s,\sigma}a^{\dagger}_{s,\sigma}a_{s,\sigma}\quad\text{and}\quad H_D = \sum_{d,\sigma}\varepsilon_{d,\sigma}a^{\dagger}_{d,\sigma}a_{d,\sigma}. 
\end{equation}
The operators $a_{s/d,\sigma}^{\dagger}(a_{s/d,\sigma})$ represent creation(annihilation) of an electron in the single-particle state $s/d$ with spin $\sigma$ and free energy $\varepsilon_{s/d}$.

We examine transport through an Anderson impurity, which is described by the Hamiltonian 
\begin{equation}
H_M = \sum_{\sigma}\varepsilon_{\sigma} a_{\sigma}^{\dagger}a_{\sigma}+U a^{\dagger}_{\uparrow}a_{\uparrow}a^{\dagger}_{\downarrow}a_{\downarrow},
\end{equation}
where the operator $a_{\sigma}^{\dagger}$($a_{\sigma}$) creates (annihilates) an electron with spin $\sigma$ on the single particle level with energy $\varepsilon_{\sigma}$, and $U$ is the Coulomb repulsion. When $U\rightarrow\infty$ and there are no spin split energy levels, the system can be modelled by a SRL:
\begin{equation}
H_M = \varepsilon a^{\dagger}a.
\end{equation} 

The interaction between the nanoscale quantum system and the macroscopic electrodes is described by the Hamiltonian 
\begin{equation}
\label{V}
H_T = t_S\sum_{s,\sigma}(a^{\dagger}_{s,\sigma}a_{\sigma}+a_{\sigma}^{\dagger}a_{s,\sigma})+t_D\sum_{d,\sigma}(a^{\dagger}_{d,\sigma}a_{\sigma}+a_{\sigma}^{\dagger}a_{d,\sigma}),
\end{equation}
where $t_S$ and $t_D$ are tunneling amplitudes between the molecule and source and drain electrode, respectively.

The quantum system has four states; it can either be empty ($\langle0|$), occupied by a single spin up electron ($\langle\uparrow|$), occupied by a single spin down electron ($\langle\downarrow|$), or occupied by a spin up and spin down electron ($\langle2|$). These states have the associated probabilities $P_{0}=\bra{0}\rho\ket{0}$, $P_{\uparrow}=\bra{\uparrow}\rho\ket{\uparrow}$, $P_{\downarrow}=\bra{\downarrow}\rho\ket{\downarrow}$, and $P_{2}=\bra{2}\rho\ket{2}$, where $\rho$ is the reduced density matrix of the Anderson impurity. 

The dynamics of the system is defined by a quantum master equation, which is constructed from quantum rates associated with electron tunneling processes. The rate of transforming from reduced system state $m$ to reduced system state $n$ is denoted $\Gamma_{nm}$.
To calculate the $\Gamma_{nm}$ we use the T-matrix approach, which is suitable as it provides a direct method for calculating transition rates between eigenstates of quantum many-body systems. Cotunneling has previously been explored via a comprehensive real-time diagrammatic method;\cite{Schoeller1994,Konig1997,Thielmann2005,Braggio2006} however, the T-matrix approach is a suitable approximation for this more rigorous method when the dynamics does not exhibit non-Markovian phenomena.\cite{PhysRevLett.95.146806} This occurs for large temperatures $k_{B}T\gg\gamma$ alongside a large gap between the Fermi energies of the baths and the energy levels participating in the transport $\delta=|\text{eV}-\varepsilon|\gg\gamma$, a condition that is met in the Coulomb blockade regime. Our calculations are all performed with $\gamma=0.5k_{B}T$, which falls within this regime. Furthermore, the construction of the WTD requires that back-scattering from the drain is not included, and so in the tunneling regime we necessarily have $\gamma\ll k_{B}T\ll\delta$. Finally, using the rate equation requires the secular approximation; coherences in the off-diagonals of the full density matrix are ignored. 

For the sake of self-completeness and to introduce relevant notations used throughout the paper, below we explicitly derive the sequential and cotunneling rates used in the quantum master equation. Here, we briefly summarise the method outlined by Bruus and Flensberg.\cite{Bruus2002,PhysRevB.77.195416} First, the Hamiltonian is reformulated as 
\begin{equation}
H(t) = H_{S}+H_{D}+H_{M}+H_Te^{\eta t},
\end{equation}
where the time-independent part $H_0 = H_S + H_D + H_M$ has a trivial but fast time-evolution $e^{-iH_{0}t}$, and the complex but slow time-evolution is due to the interaction $H_Te^{\eta t}$, which is treated as a perturbation. The time factor $e^{\eta t}$ ensures that the perturbation is turned on adiabatically at $t=-\infty$ by assuming that $\eta$ is an infinitesimal positive real number.

The starting point for the T-matrix approach is the probability $P_{f}(t)$ that the system is in state $\ket{f}$ at time $t$ given that time $t=0$ it was in state $\ket{i}$, which is just the square of their overlap; and from here, the transition rate between the two states is the time derivative of $P_{f}(t)$:
\begin{equation}
\label{rate}
\Gamma_{fi} = \frac{d}{dt}\big{|}\braket{f|i(t)}\big{|}^2.
\end{equation}
Using the interaction picture, Eq.\eqref{rate} is transformed to 
\begin{equation}
\label{General FGR}
\Gamma_{fi} = 2\pi \big{|}\bra{f}T\ket{i}\big{|}^2\delta(E_{i}-E_{f}),
\end{equation} where the T-matrix is 
\begin{align}
T & = H_T+H_T\frac{1}{E_{i}-H_{0}+i\eta}H_T \nonumber \\
	& +H_T\frac{1}{E_{i}-H_{0}+i\eta}H_T\frac{1}{E_{i}-H_{0}+i\eta}H_T+... .
\end{align}

The sequential tunneling regime corresponds to second order in $H_T$ in the transition rates: the first linear term in the $T$-matrix. So the sequential rates are 
\begin{equation}
\Gamma_{fi} = 2\pi \big{|}\bra{f}H_T\ket{i}\big{|}^2\delta(E_{i}-E_{f}),
\end{equation} which is just the standard Fermi's Golden Rule. 

In the many-body  configuration in the full Fock space, the initial and final states are tensor products of the discrete system states (molecular or quantum dot) and continuous electrode states: $\ket{m}\otimes\ket{i_{S/D}}$ and $\ket{n}\otimes\ket{f_{S/D}}$, with eigen-energies $E_m+\varepsilon_{i_{s/d}}$ and $E_n+\varepsilon_{f_{s/d}}$ respectively. Consequently, there are multiple final and initial states that correspond to a system state of $\ket{m/n}$; they must be summed over, and the initial states weighted with a thermal distribution function $W^{S/D}_{i_{m}}$: 
\begin{align}
\Gamma_{nm}^{S/D} & = 2\pi\sum_{f_{S/D},i_{S/D}}\big{|}\bra{f_{S/D}}\bra{n}H_T^{S/D}\ket{m}\ket{i_{S/D}}\big{|}^{2}W^{S/D}_{i_{m}} \nonumber \\
 & \times\delta(E_{m}-E_{n}+\varepsilon_{i_{S/D}}-\varepsilon_{f_{S/D}}).
\end{align}
At this point we can now calculate the sequential rates for electron tunneling between the electrodes and the system:
\begin{align} 
\Gamma_{\sigma0}^{S/D} & = \gamma^{S/D}n_{F}(\varepsilon_{\sigma}-\mu_{S/D}), \\
\Gamma_{0\sigma}^{S/D} & = \gamma^{S/D}\big{(}1-n_{F}(\varepsilon_{\sigma}-\mu_{S/D})\big{)}, \\
\Gamma_{\sigma2}^{S/D} & = \gamma^{S/D}\big{(}1-n_{F}(\varepsilon_{\sigma}+U-\mu_{S/D})\big{)}, & \text{and} \\
\Gamma_{2\sigma}^{S/D} & = \gamma^{S/D}(\varepsilon_{\sigma}+U-\mu_{S/D}), \\
\nonumber\end{align}
where $\gamma^{S/D}=2\pi\big{|}t_{S/D}\big{|}^2\rho(\varepsilon_{S/D})$ and $\rho(\varepsilon_{S/D})$ is the density of states for the source and drain electrodes, which is assumed to be constant. In the limiting case of a SRL, the rates reduce to 
\begin{align}
\Gamma_{10}^{S/D} & = \gamma^{S/D}n_{F}(\varepsilon-\mu_{S/D}) & \text{and} \\
\Gamma_{01}^{S/D} & = \gamma^{S/D}\big{(}1-n_{F}(\varepsilon-\mu_{S/D})\big{)}.
\end{align}
Throughout the paper we use a symmetric coupling, such that $\gamma^{S}=\gamma^{D}=\frac{\gamma}{2}$. The $n_{F}(\varepsilon-\mu_{S/D})$ are the Fermi-Dirac distributions for the source and drain electrodes: 
\begin{equation}
n_{F}(\varepsilon-\mu_{S/D})  = \frac{1}{1+e^{(\varepsilon-\mu_{S/D})\beta}},
\end{equation}
where $\beta=\frac{1}{k_{B}T}$. When the electronic level is within the bias window and in the limit of infinite source-drain bias, which is achieved by making the voltage $\mu_{S}-\mu_{D}$ large, the configuration undergoes forward tunneling only: that is, from the source to the molecule or from the molecule to the drain. However, in the Coulomb blockade regime the electronic levels are outside the bias window, regardless of the large voltage. To reconcile the two scenarios we note that their combined processes are tunneling from the source to the molecule, from the molecule to the source, and from the molecule to the drain. 
In effect, the total sequential rates for an Anderson impurity reduce to $\Gamma_{\sigma0}=\Gamma_{\sigma0}^{S}$, $\Gamma_{0\sigma}=\Gamma_{0\sigma}^{S}+\Gamma_{0\sigma}^{D}$, $\Gamma_{2\sigma}=\Gamma_{2\sigma}^{S}$, and $\Gamma_{\sigma2}=\Gamma_{\sigma2}^{S}+\Gamma_{\sigma2}^{D}$. Similarly, the total sequential rates for a SRL are $\Gamma_{10}=\Gamma_{10}^{S}$ and $\Gamma_{01}=\Gamma_{01}^{S}+\Gamma^{D}$, where we have adopted the shorthand $\Gamma^{D}=\Gamma_{01}^{D}$.

The next-to-leading term in the T-matrix expansion is second order in the tunneling coupling $\gamma$, which is fourth order in $H_T$ in the rate expression, and describes cotunneling effects. For an Anderson impurity in the infinite bias limit there are multiple cotunneling pathways, which can be categorised as either inelastic or elastic. 

Elastic cotunneling processes leave the system in the same energetic state; for example, an electron tunnels into an empty system from the source and another electron tunnels out to the drain in the same process, leaving the molecule empty and with an extra electron in the drain. We denote the transition rate of this process $\Gamma^{SD}_{00}$, where $SD$ implies that an electron is moved from the source to the drain. Note that this process can occur for $\uparrow$ or $\downarrow$ electrons, so that there are actually two pathways contained within the rate $\Gamma^{SD}_{00}$. Similarly, we define $\Gamma^{SD}_{22}$ as the rate of elastically tunneling through an originally doubly occupied system from the source to the drain. In this scenario the first process must be an electron tunneling from the molecule to the drain, which is then replaced by an electron from the source. Again, the process can occur for either $\uparrow$ or $\downarrow$ electrons, so the rate contains contributions from both pathways. Finally, we define $\Gamma^{SD}_{\sigma\sigma}$ as the rate of elastically cotunneling from the source to the drain through an originally $\sigma$ occupied system. This process can occur via an original tunneling of a $\bar{\sigma}$ electron from the source to the molecule followed by a subsequent tunneling of a $\bar{\sigma}$ electron from the molecule to the drain, or by the $\sigma$ electron tunneling to the drain first; so it too has two contributions to the rate. Since the system is experiencing infinite bias voltage, cotunneling processes that move an electron from the drain to the source do not contribute to the transport. 

Inelastic cotunneling processes leave the system occupied by the same number of electrons, but in a different energy state. For an Anderson impurity the only inelastic cotunneling processes are those that transform the system from being occupied by a single $\sigma$ electron to being occupied by a single $\bar{\sigma}$ electron. The rate of moving an electron from the source to the drain and changing the system occupatipon from $\sigma$ from $\bar{\sigma}$ is then $\Gamma^{SD}_{\bar{\sigma}\sigma}$. This rate has two processes as well; either a $\sigma$ electron tunnels from the molecule to the drain and is replaced from the source by a $\bar{\sigma}$ electron, or a $\bar{\sigma}$ electron tunnels from the source to the molecule and then a $\sigma$ electron tunnels from the molecule to the drain. We also define inelastic cotunneling processes involving the same electrode: $\Gamma^{SS}_{\bar{\sigma}\sigma}$ and $\Gamma^{DD}_{\bar{\sigma}\sigma}$. Although these processes do not move electrons across the system, they affect the occupation probabilities of the impurity and thus are included in the transport description. 

The wide variety of cotunneling rates involved in transport through an Anderson impurity are all derived by going to fourth order in $H_T$, so that Eq.\eqref{General FGR} becomes
\begin{widetext}
\begin{equation}
\label{Cotunneling general}
\Gamma_{n'n}^{\alpha\beta} = 2\pi\lim_{\eta\rightarrow 0^{+}}\sum_{\alpha,\beta=S,D}\sum_{i,f}\big{|}\bra{f}\bra{n'}H_T^{\beta}\frac{1}{E_{i,n}-H_{0}+i\eta}H_T^{\alpha}\ket{n}\ket{i}\big{|}^{2} W_{i,n}^{\alpha}W_{i,n}^{\beta}\times\delta(\varepsilon_{i}-\varepsilon_{f}),
\end{equation}
\end{widetext}
where $n=n'$ for elastic cotunneling processes and $n\neq n'$ for inelastic cotunneling processes, and the notation recognises the fact that cotunneling always leaves the system occupied by the same number of electrons as it was before the process.

It is assumed that the thermal probabilities for the source and drain are independent and so can be factored: $W^{S}_{i_{n}}W^{D}_{i_{n}}$. Additionally, we assume weak coupling, so that the electrode thermal probabilites are independent of the state of the quantum system at time $t=t_{0}$. The imaginary component $i\eta$ in Eq.\eqref{Cotunneling general} ensures that, due to tunneling processes not included in a second-order expansion, the intermediate energy of the intermediate virtual state has a finite width, and with its inclusion divergent integrals in the rate are avoided. The inclusion of $i\eta$, and the assumption that it is $\mathcal{O}(\gamma)$, forms the first part of a regularisation procedure necessary to calculate cotunneling rates. The second component of regularisation is removing any parts of the rate that are $\mathcal{O}(\gamma)$, as they correspond to a sequential tunneling. These appear because any cotunneling process can also be achieved via two sequential tunneling processes. The regularisation procedure we follow is that detailed by Koch et al.,\cite{Koch2004,Koch2006} which is equivalent to the method outlined by Turek and Matveev,\cite{Turek2002} where the finite energy width was first noted by Averin.\cite{Averin1994} Evaluating Eq.\eqref{Cotunneling general}, applying the regularisation procedure, and taking the appropriate limits, one obtains the general form of the elastic cotunneling rates for an Anderson impurity as 
\begin{widetext}
\begin{align} \label{Cotunneling Elastic Final Rate}
\Gamma_{nn}^{SD} & = \gamma^{S}\gamma^{D} n_{B}(\mu_{D}-\mu_{S}) 
\Bigg{[}\frac{\beta}{4\pi^{2}}\Im\Bigg{\{}\psi^{(1)}\Big{(}\frac{1}{2}+\frac{i \beta}{2\pi}(\mu_{D}-E_{1})\Big{)} - \psi^{(1)}\Big{(}\frac{1}{2}+\frac{i \beta}{2\pi}(\mu_{S}-E_{1})\Big{)} \nonumber \\
& + \psi^{(1)}\Big{(}\frac{1}{2}+\frac{i \beta}{2\pi}(\mu_{D}-E_{2})\Big{)} - \psi^{(1)}\Big{(}\frac{1}{2}+\frac{i \beta}{2\pi}(\mu_{S}-E_{2})\Big{)}\Bigg{\}} \pm \frac{1}{\pi(E_{1}-E_{2})}\Re\Bigg{\{}\psi\Big{(}\frac{1}{2}-\frac{i \beta}{2\pi}(\mu_{S}-E_{2})\Big{)} \nonumber \\
&  - \psi\Big{(}\frac{1}{2}-\frac{i \beta}{2\pi}(\mu_{S}-E_{1})\Big{)} - \psi\Big{(}\frac{1}{2}-\frac{i \beta}{2\pi}(\mu_{D}-E_{2})\Big{)} + \psi\Big{(}\frac{1}{2}-\frac{i \beta}{2\pi}(\mu_{D}-E_{1})\Big{)}\Bigg{\}}\Bigg{]}, \nonumber \\
\end{align}
\end{widetext}
where $E_{1}$ and $E_{2}$ refer to the energies of the tunneling pathways involved in the process and the $\pm$ is negative only for $\Gamma_{\sigma\sigma}^{SD}$. Furthermore, the transition rates defined in Eq.\eqref{Cotunneling Elastic Final Rate} use the digamma $\psi(x)$ and trigamma $\psi^{(1)}(x)$ functions, as well as the Bose-Einstein distribution function $n_{B}(\mu_{D}-\mu_{S})$: 
\begin{equation}
n_{B}(\mu_{D}-\mu_{S}) = \frac{1}{e^{(\mu_{D}-\mu_{S})\beta}-1}.
\end{equation} 

The inelastic cotunneling rates are similarly defined :

\begin{widetext}
\begin{align} \label{Cotunneling Inelastic Final Rate}
\Gamma_{\bar{\sigma}\sigma}^{\alpha\beta} & = \gamma^{\alpha}\gamma^{\beta} n_{B}(\mu_{\beta}-\mu_{\alpha}-\varepsilon_{\sigma}+\varepsilon_{\bar{\sigma}}) 
\Bigg{[}\frac{\beta}{4\pi^{2}}\Im\Bigg{\{}\psi^{(1)}\Big{(}\frac{1}{2}+\frac{i \beta}{2\pi}(\mu_{\beta}-(\varepsilon_{\sigma}+U))\Big{)} - \psi^{(1)}\Big{(}\frac{1}{2}+\frac{i \beta}{2\pi}(\mu_{\alpha}-(\varepsilon_{\bar{\sigma}}+U))\Big{)} \nonumber \\
& + \psi^{(1)}\Big{(}\frac{1}{2}+\frac{i \beta}{2\pi}(\mu_{\beta}-\varepsilon_{\sigma})\Big{)} - \psi^{(1)}\Big{(}\frac{1}{2}+\frac{i \beta}{2\pi}(\mu_{\alpha}-\varepsilon_{\bar{\sigma}})\Big{)}\Bigg{\}} - \frac{1}{\pi U}\Re\Bigg{\{}\psi\Big{(}\frac{1}{2}-\frac{i \beta}{2\pi}(\mu_{\alpha}-\varepsilon_{\bar{\sigma}})\Big{)} \nonumber \\
& - \psi\Big{(}\frac{1}{2}-\frac{i \beta}{2\pi}(\mu_{\alpha}-(\varepsilon_{\bar{\sigma}}+U))\Big{)} - \psi\Big{(}\frac{1}{2}-\frac{i \beta}{2\pi}(\mu_{\beta}-\varepsilon_{\sigma})\Big{)} + \psi\Big{(}\frac{1}{2}-\frac{i \beta}{2\pi}(\mu_{\beta}-(\varepsilon_{\sigma}+U))\Big{)}\Bigg{\}}\Bigg{]}. \nonumber \\
\end{align}
\end{widetext}

For a SRL the number of cotunneling processes is much more limited; either an electron tunnels into the empty level from the source and another electron tunnels out to the drain in the same quantum process, or an electron tunnels out from the level into the drain and is replaced by an electron from the source in the same quantum process. The two processes have transition rates $\Gamma^{(2)}_{00}$ and $\Gamma^{(2)}_{11}$ respectively, and one can show that $\Gamma^{(2)}_{00}=\Gamma^{(2)}_{11}=\Gamma^{(2)}$. Since the same molecular energy level is filled and emptied, both processes are elastic: 
\begin{align} \label{Cotunneling SRL final rate}
\Gamma^{(2)} & =\beta\frac{\Gamma^{S}\Gamma^{D}}{4\pi^{2}}n_{B}(\mu_{D}-\mu_{S}) \nonumber \\
&\times\Im\Bigg{\{}\psi^{(1)}\Big{(}\frac{1}{2}+\frac{i \beta}{2\pi}(\varepsilon-\mu_{S}\Big{)} - \psi^{(1)}\Big{(}\frac{1}{2}+\frac{i \beta}{2\pi}(\varepsilon-\mu_{D}\Big{)}\Bigg{\}}.
\end{align}

The details of the derivations for Eq.\eqref{Cotunneling Elastic Final Rate}, Eq.\eqref{Cotunneling Inelastic Final Rate}, and Eq.\eqref{Cotunneling SRL final rate} are in Appendix A.

From here it is tempting to construct the standard rate equation for occupation probabilities of the impurity. However, since elastic cotunneling rates do not change the state of the quantum system, they do not contribute to the rate equation for the system state probabilities. Instead, one must consider the $n$-resolved system state probabilities.
\subsection{$n$-resolved master equation}
The master equation can be resolved upon the number of electrons transferred to the drain; so $P_{0}(n,t)$ is the probability that the system is empty at time $t$ and that $n$ electrons were transferred to the drain in the interval $[0,t]$, and similarly for $P_{\sigma}(n,t)$ and $P_{2}(n,t)$. For the infinite bias regime $n=0,1,2,3, ... ,+\infty$. Thus the total probability that $n$ electrons were transferred by time $t$ is 
\begin{align}
P(n,t) & = (\mathbf{I},\mathbf{P}(n,t)) \\
& = P_{0}(n,t)+P_{\uparrow}(n,t)+P_{\downarrow}(n,t)+P_{2}(n,t),
\end{align}
where $\mathbf{I}$ is the identity vector
\begin{align}
\mathbf{I} & =
\begin{bmatrix}
1 & 1 & 1 & 1
\end{bmatrix}
\end{align}
and $\mathbf{P}(n,t)$ is the probability vector
\begin{align}
\mathbf{P}(n,t) & = 
\begin{bmatrix}
P_{0}(n,t) \\
P_{\uparrow}(n,t) \\
P_{\downarrow}(n,t) \\
P_{2}(n,t)
\end{bmatrix}.
\end{align} 

The $n$-resolved Markovian master equation follows the  general form 
\begin{equation}
\dot{\mathbf{P}}(n,t)=\sum_{n'}\mathbf{L}(n-n')\mathbf{P}(n,t).
\end{equation}
For the tunneling interaction defined in Eq.(\ref{V}) each $n$ is connected only to its neighbouring values $n'=n,n\pm1$ and for an Anderson impurity in the infinite bias regime, including cotunneling processes, the $n$-resolved master equation is intuitively 
\begin{widetext}
\begin{align} \label{nresolved}
\dot{\mathbf{P}}(n,t) & = \left[\begin{array}{cccc}
 -(\Gamma_{\uparrow0}^{S}+\Gamma_{\downarrow0}^{S} +\Gamma_{00}^{SD}) 
 & \Gamma_{0\uparrow}^{S} & \Gamma_{0\downarrow}^{S} & 0 \\
\\
\Gamma_{\uparrow 0}^{S} & -(\Gamma_{0\uparrow}+\Gamma_{2\uparrow}^{S} + \Gamma_{\uparrow\uparrow}^{SD}+\Gamma_{\downarrow\uparrow}^{(2)}) & \Gamma_{\uparrow\downarrow}^{SS}+\Gamma_{\uparrow\downarrow}^{DD} & \Gamma_{\uparrow2}^{S}\\
\\
\Gamma_{\downarrow0}^{S} & \Gamma_{\downarrow\uparrow}^{SS}+\Gamma_{\downarrow\uparrow}^{DD} &  -(\Gamma_{0\downarrow}+\Gamma_{2\downarrow}^{S} + \Gamma_{\downarrow\downarrow}^{SD} +  \Gamma_{\uparrow\downarrow}^{(2)}) & \Gamma_{\downarrow2}^{S}\\
\\
0 & \Gamma_{2\uparrow}^{S} & \Gamma_{2\downarrow}^{S} &  -(\Gamma_{\uparrow2}+\Gamma_{\downarrow2} +\Gamma_{22}^{SD})
\end{array}\right] \mathbf{P}(n,t) \nonumber \\
& \nonumber \\
& + \left[\begin{array}{cccc}
\Gamma_{00}^{SD} & \Gamma_{0\uparrow}^{D} & \Gamma_{0\downarrow}^{D} & 0\\
& \\
0 & \Gamma_{\uparrow\uparrow}^{SD} & \Gamma_{\uparrow\downarrow}^{SD} & \Gamma_{\uparrow2}^{D}\\
& \\
0 & \Gamma_{\downarrow\uparrow}^{SD} & \Gamma_{\downarrow\downarrow}^{SD} & \Gamma_{\downarrow2}^{D}\\
& \\
0 & 0 & 0 & \Gamma_{22}^{SD}
\end{array}\right] \mathbf{P}(n-1,t).
\end{align}
\end{widetext}
Here, we have excluded those rates that involve back-tunneling processes from the drain, as they have a negligible contribution in the infinite bias regime. Additionally, we use the notation
\begin{align}
\Gamma^{(2)}_{\bar{\sigma}\sigma} & =\Gamma_{\bar{\sigma}\sigma}^{SS}+\Gamma_{\bar{\sigma}\sigma}^{DD}+\Gamma_{\bar{\sigma}\sigma}^{SD}.
\end{align}

Evidently, the $n$-resolved master equation is an infinite set of coupled equations since $n=0,1,2,...,+\infty$. To solve, we use the elegant idea,  proposed first by Nazarov and extended to master equations by Bagrets and Nazarov, of introducing a continuous counting field $\chi$, with $0\geq\chi\geq2\pi$:\cite{Nazarov1999,Bagrets2003}
\begin{align}
\mathbf{P}(\chi,t)&=\sum_{n}e^{in\chi}\mathbf{P}(n,t), & \text{and} \\
\mathbf{P}(n,t)&=\frac{1}{2\pi}\int_{0}^{2\pi}e^{-in\chi}\mathbf{P}(\chi,t)d\chi.
\end{align}
Multiplying Eq.\eqref{nresolved} by $e^{in\chi}$ and transforming $\sum_{n}e^{in\chi}\mathbf{P}(n-1,t) \rightarrow \sum_{m}e^{i(m+1)\chi}\mathbf{P}(m,t)$, one obtains the $n$-resolved master in $\chi$-space in the form $\mathbf{\dot{P}}(\chi,t)=\mathbf{L}(\chi)\mathbf{P}(\chi,t)$:
\begin{widetext}
\begin{align} \label{chiME}
\frac{d}{dt}\left[\begin{array}{c} P_{0}(\chi,t) \\ P_{\uparrow}(\chi,t) \\ P_{\downarrow}(\chi,t) \\  P_{2}(\chi,t) \end{array}\right]
 & = \left[\begin{array}{cccc}
\begin{array}{@{}c@{}} -(\Gamma_{\uparrow0}^{S}+\Gamma_{\downarrow0}^{S}) \\ +\Gamma_{00}^{SD}(e^{i\chi}-1) \end{array}
 & \Gamma_{0\uparrow}^{S}+\Gamma_{0\uparrow}^{D}e^{i\chi} & \Gamma_{0\downarrow}^{S}+\Gamma_{0\downarrow}^{D}e^{i\chi} & 0 \\
\\
\Gamma_{\uparrow 0}^{S} & \begin{array}{@{}c@{}} -(\Gamma_{0\uparrow}+\Gamma_{2\uparrow}^{S}+\Gamma_{\downarrow\uparrow}^{(2)}) \\ +\Gamma_{\uparrow\uparrow}^{SD}(e^{i\chi}-1) \end{array} & \Gamma_{\uparrow\downarrow}^{SS}+\Gamma_{\uparrow\downarrow}^{DD}+\Gamma_{\uparrow\downarrow}^{SD}e^{i\chi} & \Gamma_{\uparrow2}^{S}+\Gamma_{\uparrow2}^{D}e^{i\chi}\\
\\
\Gamma_{\downarrow0}^{S} & \Gamma_{\downarrow\uparrow}^{SS}+\Gamma_{\downarrow\uparrow}^{DD}+\Gamma_{\downarrow\uparrow}^{SD}e^{i\chi} & \begin{array}{@{}c@{}} -(\Gamma_{0\downarrow}+\Gamma_{2\downarrow}^{S} + \Gamma_{\uparrow\downarrow}^{(2)}) \\ + \Gamma_{\downarrow\downarrow}^{SD}(e^{i\chi}-1) \end{array} & \Gamma_{\downarrow2}^{S}+\Gamma_{\downarrow2}^{D}e^{i\chi}\\
\\
0 & \Gamma_{2\uparrow}^{S} & \Gamma_{2\downarrow}^{S} & \begin{array}{@{}c@{}} -(\Gamma_{\uparrow2}+\Gamma_{\downarrow2}) \\ +\Gamma_{22}^{SD}(e^{i\chi}-1)\end{array}
\end{array}\right] \left[\begin{array}{c} P_{0}(\chi,t) \\ P_{\uparrow}(\chi,t) \\ P_{\downarrow}(\chi,t) \\  P_{2}(\chi,t) \end{array}\right].
\end{align}
\end{widetext}

Eq.\eqref{chiME} has the formal solution:
\begin{equation}
\mathbf{P}(\chi,t)=e^{\mathbf{L}(\chi)t}\mathbf{P}(\chi,0),
\end{equation}
where the inital condition is $\mathbf{P}(\chi,0)=\mathbf{P}(n=0,0)$, since it is assumed that electron counts are monitored after $t=0$.
We also assume that the system was prepared in the steady state at $t=0$, so that $\mathbf{P}(n=0,0)=\bar{\mathbf{P}}$ with $\bar{\mathbf{P}}$ being a null vector of the standard Liouvillian:
\begin{equation}
\mathbf{L}(0) \bar{\mathbf{P}}=0.
\end{equation}

Then, the probability that $n$ electrons have been transferred to the drain by time $t$ is 
\begin{equation}
P(n,t)=\frac{1}{2\pi}\int_{0}^{2\pi}e^{-in\chi}(\mathbf{I},e^{\mathbf{L}(\chi)t}\bar{\mathbf{P}})d\chi.
\end{equation}
At this point one could define a moment-generating function $M(\chi,t)=(\mathbf{I},e^{\mathbf{L}(\chi)t}\bar{\mathbf{P}})$ and derive the moments of transferred charge $\braket{n^{k}}=(-i)^{k}\frac{\partial^{k}}{\partial\chi^{k}}M(\chi,t)\Big{|}_{\chi=0}$ to obtain the full counting statistics. However, we are interested in the WTD. 

%%%%%%%%%%%%%%%%%%%%%%%%%%%%%%%%%%%%%%%%%%%%%%%%%%%%%%%%%%%%%%%%%%%%%%%%%%%%%%%%%%%%%%%
\subsection{WTD definition}
Based on the ideas from quantum optics single photon counting theories,\cite{Srinivas2010,PhysRevA.39.1200} Brandes first introduced the concept of a WTD to electron transport with a formalism that used ``jump'' operators defined from the master equation of the system.\cite{Brandes2008} In order to include cotunneling rates, however, we will start with the conditional WTD defined in terms of the idle time probability:\cite{VanKampen1981,Thomas2013}
\begin{equation} \label{WTD_from_IT}
w(\tau)=\frac{1}{p}\frac{\partial^{2}}{\partial\tau^{2}}\Pi({\tau}).
\end{equation}
The idle time probability $\Pi(\tau)$ is the probability that there were no electron tunnelings to the drain in the measurement time $\tau$. Here, $p$ is the initial probability of observing an electron tunneling to the drain, and can be defined in terms of $\Pi(\tau)$ as well: $p=-\frac{\partial}{\partial\tau}\Pi(\tau)\Big{|}_{\tau=0}$.
 The key relation is that the idle time probability is the probability for no electrons to be transferred to the drain between time $t=0$ and time $t=\tau$, so that when forward tunneling only is included $\Pi(\tau)=P(0,\tau)$.\cite{Thomas2013} The moment-generating function can be written as 
\begin{equation}
M(\chi,\tau)=P(0,\tau)+\sum_{n=1}^{\infty}e^{in\chi}P(n,\tau);
\end{equation}
hence in the infinite bias regime the idle time distribution is 
\begin{equation}
\Pi(\tau)=\lim_{\chi\rightarrow i\infty} (\mathbf{I},e^{\mathbf{L}(\chi)\tau}\bar{\mathbf{P}}).
\end{equation}
Combining with the definition of the WTD from Eq \eqref{WTD_from_IT}, we get 
\begin{equation}
w(\tau) = -\lim_{\chi\rightarrow i\infty}\frac{(\mathbf{I},\mathbf{L}(\chi)e^{\mathbf{L}(\chi)\tau}\mathbf{L}(\chi)\bar{\mathbf{P}})}{(\mathbf{I},\mathbf{L}(\chi)\bar{\mathbf{P}})},
\end{equation} and in Laplace space 
\begin{equation}
\tilde{w}(z) = -\lim_{\chi\rightarrow i\infty}\frac{(\mathbf{I},\mathbf{L}(\chi)(z-\mathbf{L}(\chi))^{-1}\mathbf{L}(\chi)\bar{\mathbf{P}})}{(\mathbf{I},\mathbf{L}(\chi)\bar{\mathbf{P}})}.
\end{equation}
Similarly to the sequential tunneling case, $\mathbf{L}(\chi)$ is formally split into a quantum jump part $\mathbf{J}(\chi)=\mathbf{J}e^{i\chi}$, containing the $\chi$-dependence, and the $\chi$-independent $\mathbf{L}_{0}$:
\begin{widetext}
\begin{align}
\mathbf{L}(\chi) & = \left[\begin{array}{cccc}
 -(\Gamma_{\uparrow0}^{S}+\Gamma_{\downarrow0}^{S} +\Gamma_{00}^{SD}) 
 & \Gamma_{0\uparrow}^{S} & \Gamma_{0\downarrow}^{S} & 0 \\
\\
\Gamma_{\uparrow 0}^{S} & -(\Gamma_{0\uparrow}+\Gamma_{2\uparrow}^{S} + \Gamma_{\uparrow\uparrow}^{SD}+\Gamma_{\downarrow\uparrow}^{(2)}) & \Gamma_{\uparrow\downarrow}^{SS}+\Gamma_{\uparrow\downarrow}^{DD} & \Gamma_{\uparrow2}^{S}\\
\\
\Gamma_{\downarrow0}^{S} & \Gamma_{\downarrow\uparrow}^{SS}+\Gamma_{\downarrow\uparrow}^{DD} &  -(\Gamma_{0\downarrow}+\Gamma_{2\downarrow}^{S} + \Gamma_{\downarrow\downarrow}^{SD} +  \Gamma_{\uparrow\downarrow}^{(2)}) & \Gamma_{\downarrow2}^{S}\\
\\
0 & \Gamma_{2\uparrow}^{S} & \Gamma_{2\downarrow}^{S} &  -(\Gamma_{\uparrow2}+\Gamma_{\downarrow2} +\Gamma_{22}^{SD})
\end{array}\right] \nonumber \\
& + \left[\begin{array}{cccc}
\Gamma_{00}^{SD} & \Gamma_{0\uparrow}^{D} & \Gamma_{0\downarrow}^{D} & 0\\
& \\
0 & \Gamma_{\uparrow\uparrow}^{SD} & \Gamma_{\uparrow\downarrow}^{SD} & \Gamma_{\uparrow2}^{D}\\
& \\
0 & \Gamma_{\downarrow\uparrow}^{SD} & \Gamma_{\downarrow\downarrow}^{SD} & \Gamma_{\downarrow2}^{D}\\
& \\
0 & 0 & 0 & \Gamma_{22}^{SD}
\end{array}\right]e^{i\chi} \nonumber \\
& = \mathbf{L}_{0}+\mathbf{J}e^{i\chi}
 \label{chi_splitting}
\end{align}
\end{widetext}

The splitting is similarly defined for a single resonant level:
\begin{widetext}
\begin{align} \label{chi_splitting_SRL}
\mathbf{L}(\chi) & = \left[\begin{array}{cc}
 -(\Gamma_{10}^{S}+\Gamma^{(2)}) & \Gamma_{01}^{S} \\ 
 \Gamma_{10}^{S}& -(\Gamma^{D}+\Gamma_{01}^{S}+\Gamma^{(2)})
\end{array}\right] + \left[\begin{array}{cc}
\Gamma^{(2)} & \Gamma^{D} \\ 
0& \Gamma^{(2)}
\end{array}\right]e^{i\chi}
\end{align}
\end{widetext}

Using the splittings in Eq.\eqref{chi_splitting} and Eq.\eqref{chi_splitting_SRL} the WTD becomes 
\begin{equation}
w(\tau)  = -\lim_{\chi\rightarrow i\infty}\frac{(\mathbf{I},(\mathbf{L}_{0}+\mathbf{J}e^{i\chi})e^{(\mathbf{L}_{0}+\mathbf{J}e^{i\chi})\tau}(\mathbf{L}_{0}+\mathbf{J}e^{i\chi})\bar{\mathbf{P}})}{(\mathbf{I},(\mathbf{L}_{0}+\mathbf{J}e^{i\chi})\bar{\mathbf{P}})},
\end{equation}
which is
\begin{equation}
w(\tau)  = -\frac{(\mathbf{I},\mathbf{L}_{0} e^{\mathbf{L}_{0} \tau} \mathbf{L}_{0}\bar{\mathbf{P}})}{(\mathbf{I},\mathbf{L}_{0} \bar{\mathbf{P}})}.
\end{equation}
Noting that $\mathbf{L}_{0}=\mathbf{L}(0)-\mathbf{J}$,  $\mathbf{L}(0)\bar{\mathbf{P}}=0$ and $(\mathbf I,\mathbf{L}(0) {\mathbf{A}})=0$  for arbitrary $\mathbf A$, we obtain the standard expressions for the WTD in the time domain:
\begin{equation}
w(\tau)  = \frac{(\mathbf{I},\mathbf{J}e^{\mathbf{L}_{0}\tau}\mathbf{J}\mathbf{\bar{P}})}{(\mathbf{I},\mathbf{J}\mathbf{\bar{P}})},  \label{WTD_proper_time} 
\end{equation}
which in Laplace space becomes 
\begin{equation}
\tilde{w}(z)  = \frac{(\mathbf{I},\mathbf{J}(z-\mathbf{L}_{0})^{-1}\mathbf{J}\mathbf{\bar{P}})}{(\mathbf{I},\mathbf{J}\mathbf{\bar{P}})}. \label{WTD_proper}
\end{equation}

Here, we see that in the case of forward tunneling the WTD reduces to the one calculated using Brandes' method.\cite{Brandes2008} Despite this, the $n$-resolved master equation is still necessary as it tells us how to construct $\mathbf{L}_{0}$ from the quantum jump operator $\mathbf{J}$. 
We notice that the method breaks down if backwards tunneling processes are included, as their factor $e^{-i\chi}$ will diverge in the limit $\chi\rightarrow i\infty$. This is a serious limitation of the approach, and it is not yet clear how to resolve it.

Although the single WTD is itself an interesting quantity, in order to compute higher-order expectation values and analyse micrscopic fluctuations we must also generalise it to two or more consecutive waiting times. For example, the WTD for two waiting times, $w_2(\tau_2,\tau_1) $, is defined as the joint probability distribution that the first electron waits time $\tau_1$ and the next electron waits time $\tau_2$ before tunneling to the drain:\cite{nonrenewal-kosov,Ptaszynski2017}
\begin{align}
w_{2}(\tau_{2},\tau_{1}) & = \frac{(\mathbf{I},\mathbf{J}e^{\mathbf{L}_{0}\tau_{2}}\mathbf{J}e^{\mathbf{L}_{0}\tau_{1}}\mathbf{J}\mathbf{\bar{P}})}{(\mathbf{I},\mathbf{J}\mathbf{\bar{P}})}.
\label{wtd2-2}
\end{align}

Moments of the single WTD are easily calculable by introducing a moment-generating function over $\tau$:
\begin{equation}
\label{moment_generator}
K(x)  = \int_{0}^{\infty}d\tau e^{ix\tau}w(\tau)
 	   =  \frac{(\mathbf{I},\mathbf{J}\mathbf{G}(x_{1})\mathbf{J}\mathbf{\bar{P}})}{(\mathbf{I},\mathbf{J}\mathbf{\bar{P}})},
\end{equation}

where $x$ is a real number and 
\begin{equation}
\mathbf{G}(x) = (\mathbf{L}_{0}+ix)^{-1}.
\end{equation}
We obtain all possible moments by direct differentiation with respect to $x$, such that 
\begin{multline}
 \langle\tau^n\rangle   = \int_{0}^{\infty}d\tau\;\tau^n w(\tau)  \\
 =  n! (-1)^{n+1}\frac{(\mathbf{I},\mathbf{J}\mathbf{G}(0)^{n+1}\mathbf{J}\mathbf{\bar{P}})}{(\mathbf{I},\mathbf{J}\mathbf{\bar{P}})}.
\end{multline}

The second-order expectation value is calculated similarly:
\begin{multline}
\langle  \tau_2 \tau_1 \rangle   = \int_0^{\infty} d \tau_1 \int_0^{\infty} d \tau_2 \; \tau_1 \tau_2 w_2(\tau_2, \tau_1) \\
 = \frac{(\mathbf{I},\mathbf{J}\mathbf{G}(0)^2\mathbf{J}\mathbf{G}(0)^2\mathbf{J}\mathbf{\bar{P}})}{(\mathbf{I},\mathbf{J}\mathbf{\bar{P}})}.
\end{multline}

%%%%%%%%%%%%%%%%%%%%%%%%%%%%%%%%%%%%%%%%%%%%%%%%%%%%%%%%%%%%%%%%%%%%%%%%%%%

\section{Results} \label{Results}

In this section we analytically and numerically investigate statistics of waiting time intervals between successive electron cotunneling events, for both the SRL and an Anderson impurity.

\begin{figure*}
	\subfloat[]{\includegraphics[scale=0.5]{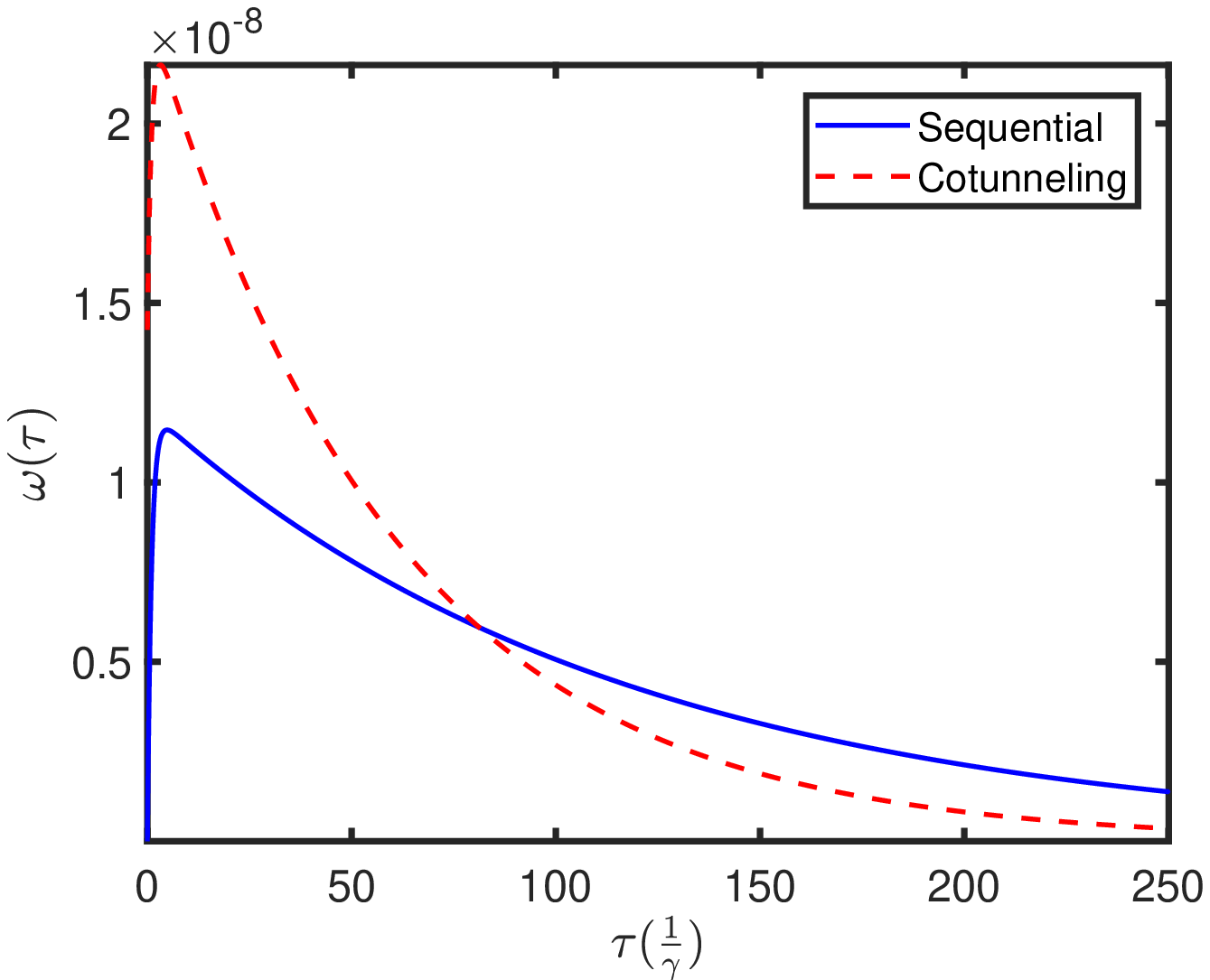}\label{WTD_CB}}
	\subfloat[]{\includegraphics[scale=0.5]{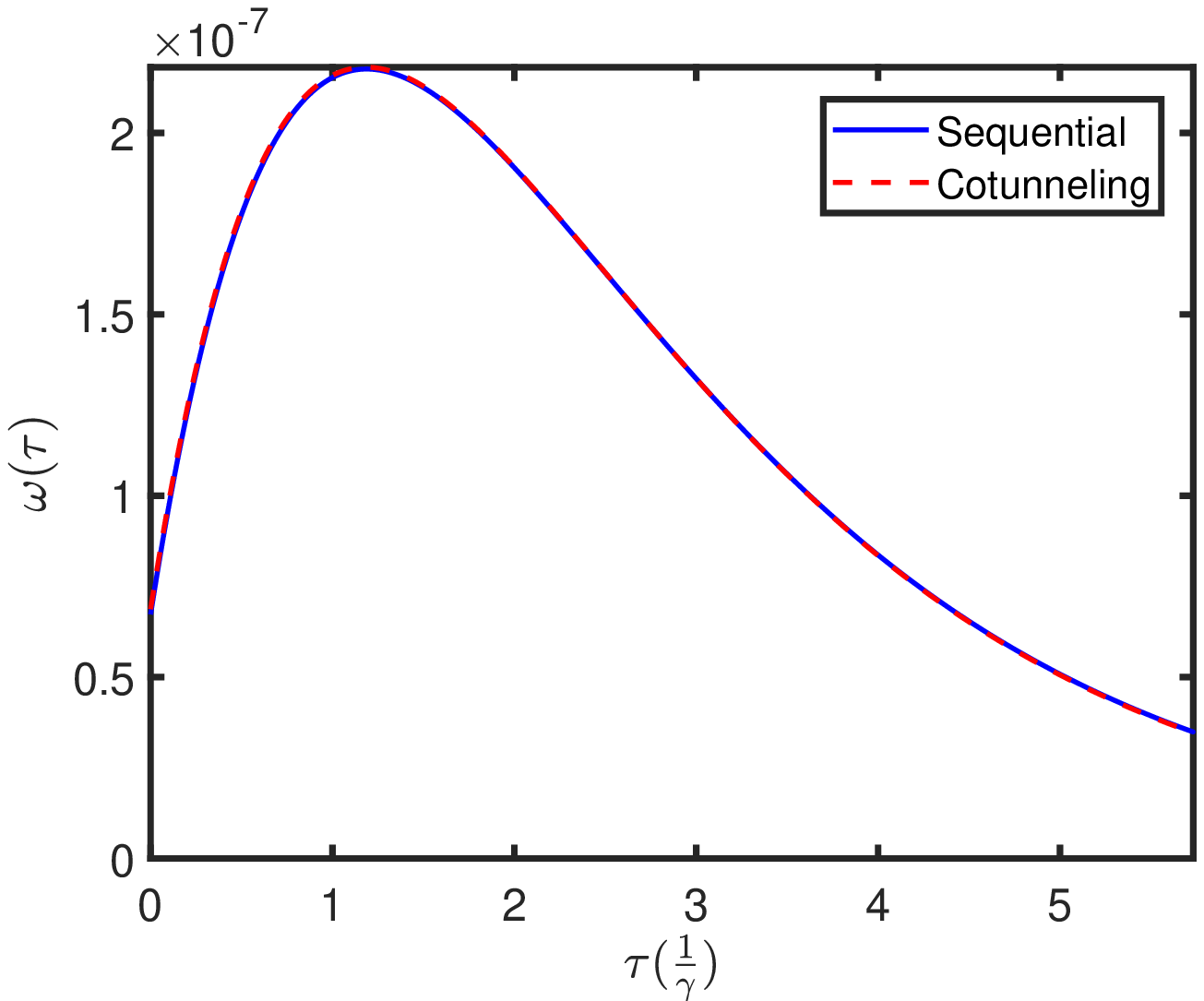}\label{WTD_Tunneling}}
    	\caption{Colour online. Sequential and cotunneling WTDs for two voltages; in (a) the level is in the Coulomb blockade regime and in (b) the level is in the tunneling regime. The y-axis represents $w_{seq}(\tau)$ and $w_{co}(\tau)$ for the sequential and cotunneling WTDs, respectively. The energies of the spin split electronic levels are $\varepsilon_{\uparrow}=0.5\text{meV}$ and $\varepsilon_{\downarrow}=1.5\text{meV}$, the Coulomb repulsion is $U=4\text{meV}$, $\text{k}_{\text{B}}\text{T}=75\mu\mbox{eV}$, and $\gamma=0.5\text{k}_{\text{B}}\text{T}$.  Parameters  for each plot are:
(a) $\mu_{S}=-\mu_{D} = 0.25\mbox{meV}$,  $\langle\tau\rangle_{seq}=2.04\mbox{ns}$,  and $\langle\tau\rangle_{co}=1.06\mbox{ns}$; 
(b) $\mu_{S}=-\mu_{D} = 5\mbox{meV}$,  $\langle\tau\rangle_{seq}=46.81\mbox{ps}$,  and $\langle\tau\rangle_{co}=46.43\mbox{ps}$.}
	\label{WTD_Time}
\end{figure*}

The WTD for a SRL in Laplace space is obtained via Eq.\eqref{WTD_proper} using the splitting from Eq.\eqref{chi_splitting_SRL}: 
\begin{equation} \label{Waiting_Time_Results_SRL_z}
\tilde{w}(z)   = \frac{a+bz}{(z+z_{+})(z+z_{-})} 
\end{equation}
and the corresponding WTD in the time domain is 
\begin{equation} \label{Waiting_Time_Results_SRL_tau}
w(\tau)  = \frac{a-bz_{-}}{z_{+}-z_{-}}e^{-z_{-}\tau}-\frac{a-bz_{+}}{z_{+}-z_{-}}e^{-z_{+}\tau},
\end{equation}
where the coefficients of the linear function in the numerator are 
\begin{multline}
a  =
\Big\{ (\Gamma^{D}\Gamma_{10}^{S})^2+\Gamma^{(2)}[\Gamma^{D}+\Gamma_{10}^{S}+\Gamma_{01}^{S}] [ (\Gamma^{(2)})^2+2\Gamma^{D}\Gamma_{10}^{S}] \\
  +(\Gamma^{(2)})^2 [ (\Gamma^{D})^2+(\Gamma_{10}^{S}+\Gamma_{01}^{S})^2+\Gamma^{D}(2\Gamma_{01}^{S}+3\Gamma_{10}^{S})]
\Big\} \\
 /\Big\{\Gamma^{D}\Gamma_{10}^{S}+\Gamma^{(2)}(\Gamma^{D}+\Gamma_{10}^{S}+\Gamma_{01}^{S})\Big\}\text{ and}
\end{multline}
\begin{equation}
b  = \frac{\Gamma^{(2)}\big{(}2\Gamma^{D}\Gamma_{10}^{S}+\Gamma^{(2)}(\Gamma^{D}+\Gamma_{10}^{S}+\Gamma_{01}^{S})\big{)}}{\Gamma^{D}\Gamma_{10}^{S}+\Gamma^{(2)}(\Gamma^{D}+\Gamma_{10}^{S}+\Gamma_{01}^{S}) }.
\end{equation}
The poles of the Laplace space WTD, which are also the exponents in the time-space WTD, are
\begin{multline}
z_{\pm}  = \frac{1}{2}\Big{(}2\Gamma^{(2)}+ \Gamma^{D}+ \Gamma_{10}^{S}+ \Gamma_{01}^{S} \\
  \pm \sqrt{(\Gamma^{D})^2+2\Gamma^{D}(\Gamma_{01}^{S}-\Gamma_{10}^{S})+(\Gamma_{01}^{S}+\Gamma_{10}^{S})^{2}})\Big{)}. 
\end{multline}
Interestingly, the position of the poles yield information on the individual source-drain couplings, similarly to the results Brandes found for sequential tunneling through a single resonant level.

The moments of the WTD can be derived analytically for a single resonant level model, using Eq.\eqref{moment_generator}. The average waiting time is 
\begin{equation}
\langle\tau\rangle=\frac{\Gamma^{D}+\Gamma_{01}^{S}+\Gamma_{10}^{S}}{\Gamma^{D}\Gamma_{10}^{S}+\Gamma^{(2)}(\Gamma^{D}+\Gamma_{10}^{S})}\\
=\frac{1}{\langle I\rangle^{(2)}},
\end{equation} where $\langle I\rangle^{(2)}$ is the forward current including cotunneling processes.

 The short time behaviour of the WTD is evident from Eq.\eqref{Waiting_Time_Results_SRL_tau}:

\begin{align}
w(0) = \frac{\Gamma^{(2)}\big{(}2\Gamma^{D}\Gamma_{10}^{S}+\Gamma^{(2)}(\Gamma^{D}+\Gamma_{10}^{S}+\Gamma_{01}^{S})\big{)}}{\Gamma^{D}\Gamma_{10}^{S}+\Gamma^{(2)}(\Gamma^{D}+\Gamma_{10}^{S}+\Gamma_{01}^{S}) }. \label{short_time_SRL}
\end{align}

For sequential tunneling only, a SRL is a single reset system; that is, after an electron tunneling to the drain the system is always left empty. Consequently, in such a regime the probability density at $\tau=0$ is zero as two electrons cannot be detected in the drain right after one another. In contrast, Eq.\eqref{short_time_SRL} shows that when cotunneling processes are included $w(0)\neq 0$, which implies that it is now a multiple reset system. Physically, this is plausible as the cotunneling processes that move electrons from the source to the drain occur regardless of the SRL occupancy.  

The short time behaviour is further characterised by the Pearson correlation coefficient:
\begin{align}
p=\frac{\braket{\tau_{1}\tau_{2}}-\braket{\tau}^{2}}{\braket{\tau^{2}}-\braket{\tau}^{2}},
\end{align}
where $\tau_{1}\text{ and }\tau_{2}$ are subsequent waiting times. For sequential tunneling through a single resonant level $p=0$, such that $w(\tau_{1},\tau_{2})=w(\tau_{1})w(\tau_{2})$ and waiting times between subsequent tunnelings to the drain are completely uncorrelated. Consequently, in such a regime the renewal assumption is satisfied. When cotunneling processes are included, however, the Pearson correlation coefficient is nonzero:

\begin{align} \label{Pearson_SRL}
p&=-\frac{A^2}{B\cdot C},
\end{align}
where the components are 
\begin{align}
A & = \Gamma^{(2)}\Gamma^{D}\Gamma_{10}^{S}, \\
B & = (\Gamma^{(2)})^2+\Gamma^{D}\Gamma_{10}^{S}+\Gamma^{(2)}(\Gamma^{D}+\Gamma_{01}^{S}+\Gamma_{10}^{S}), & \text{and}
\end{align}
\begin{multline}
C = (\Gamma^{(2)}(\Gamma^{D}+\Gamma_{01}^{S}+\Gamma_{10}^{S}))^{2}+\Gamma^{(2)}(\Gamma^{D}+\Gamma_{01}^{S}+\Gamma_{10}^{S})^{3} \\ 
+\Gamma^{D}\Gamma_{01}^{S}\Big{(}(\Gamma^{D})^{2}+2\Gamma^{D}\Gamma_{01}^{S}+(\Gamma_{01}^{S}+\Gamma_{10}^{S})^2\Big{)}.
\end{multline}

 Eq.\eqref{Pearson_SRL} shows that, contrary to sequential tunneling, electron waiting times for cotunneling through a single resonant level are negatively correlated, since $A^{2}\text{, }B\text{, and }C$ are all positive. However, the correlation is negligibly small, as expected from the small perturbative changes that cotunneling brings.

Turning now to the Anderson impurity, when neither of the spin split levels are in the voltage window, we expect the sequential current to be negligible. Consequently, in such a case, we also expect the average sequential waiting time $\langle\tau\rangle_{seq}$ to be large; that is, on average it takes a long time for electrons to be transferred from the source to the drain. 
In contrast, for an Anderson impurity experiencing Coulomb blockade, cotunneling provides a quantum pathway for electrons to tunnel through the system that is not visible in the sequential physics. This is evident in (a) of Fig.(\ref{WTD_Time}), where $\langle\tau\rangle_{co}$ is double $\langle\tau\rangle_{seq}$, whereas in the tunneling regime in (b) of Fig.(\ref{WTD_Time}) sequential processes dominate and $\langle\tau\rangle_{co}$ is comparable to $\langle\tau\rangle_{seq}$. 

At high voltages an Anderson impurity behaves as a multiple reset system, since an electron tunneling to the drain can leave the system singly occupied or empty, which is shown in Fig.(\ref{WTD_Time}b) as $w(0)\neq0$ for both sequential tunneling and cotunneling. In comparison, at low voltages double occupancy is energetically denied and it behaves as a single reset system, which is shown in Fig.(\ref{WTD_Time}a) as $w(0)=0$ for sequential tunneling. Again, however, when cotunneling processes are included the WTD displays multiple reset behaviour at short times, as cotunneling processes can leave the system either singly occupied or empty.  

\begin{figure*} 
\subfloat[]{\includegraphics[scale=0.5]{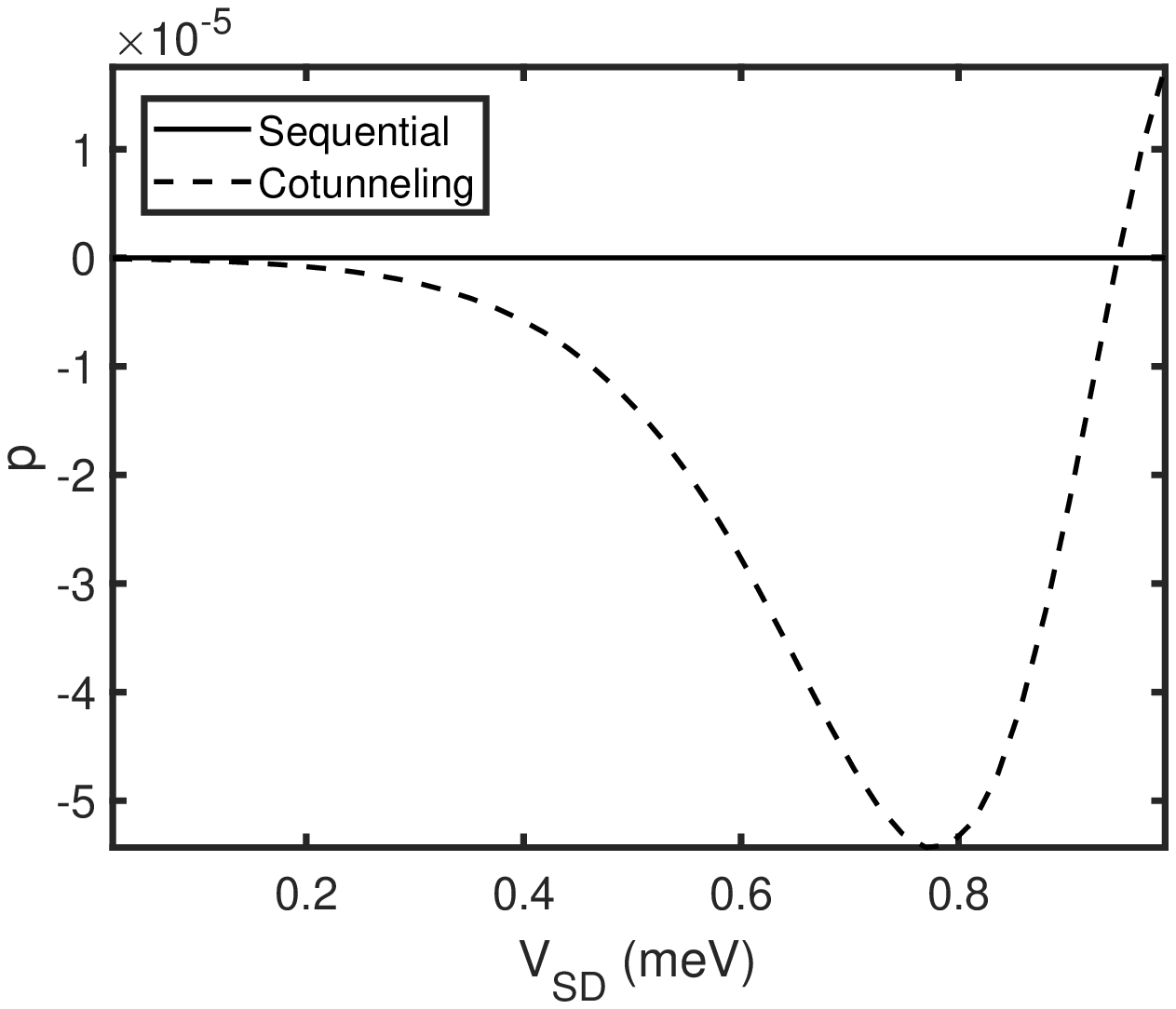}}
\subfloat[]{\includegraphics[scale=0.5]{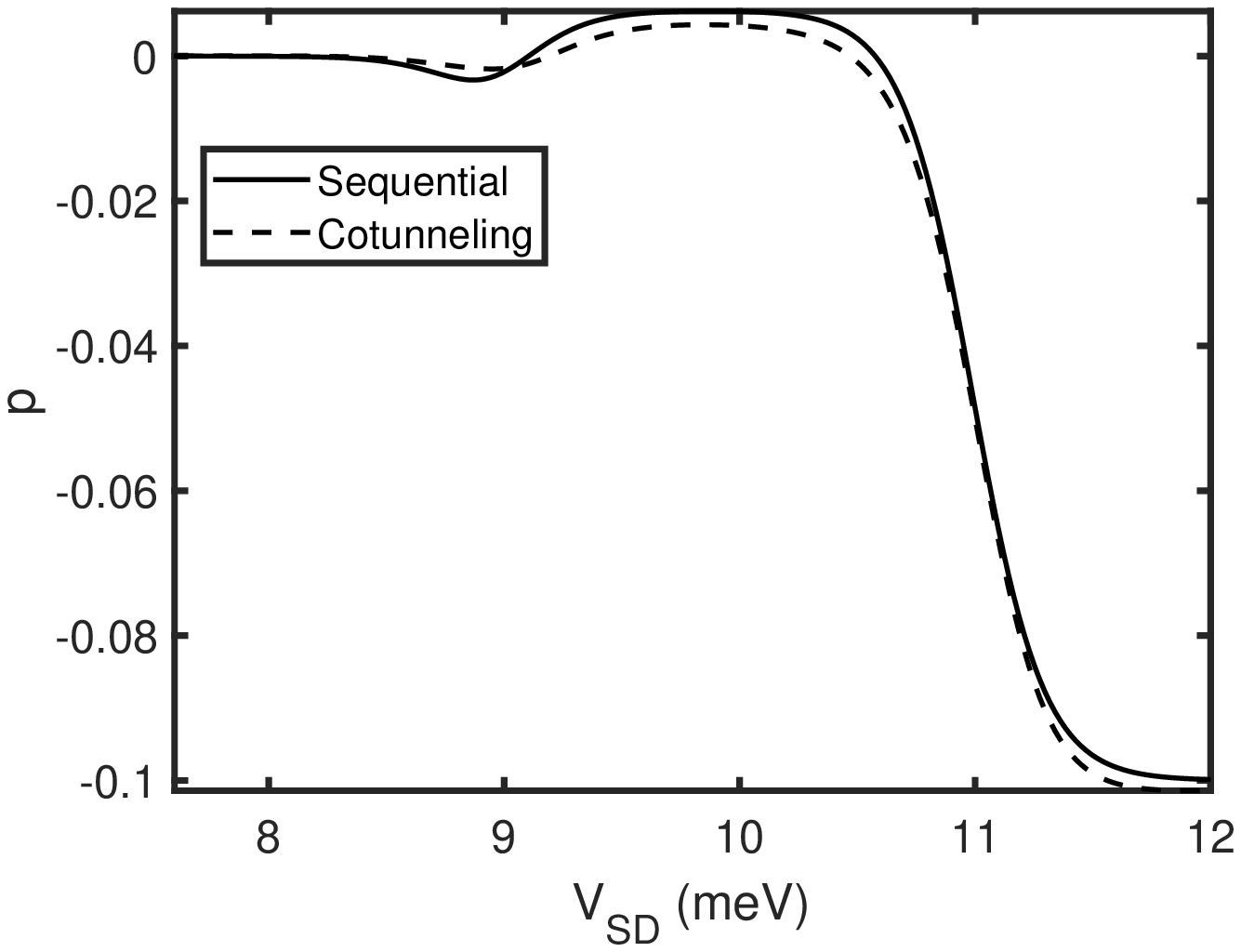}}
\caption{Pearson correlation coefficient $\frac{\langle\tau_{1}\tau_{2}\rangle-\langle\tau\rangle^{2}}{\langle\tau^{2}\rangle-\langle\tau\rangle^{2}}$ over a range of voltages in (a) the Coulomb blockade regime and (b) the tunneling regime. The energies of the spin split electronic levels are $\varepsilon_{\uparrow}=0.5\text{meV}$ and $\varepsilon_{\downarrow}=1.5\text{meV}$, the Coulomb repulsion is $U=4\text{meV}$, $\text{k}_{\text{B}}\text{T}=75\mu\mbox{eV}$, and $\gamma=0.5\text{k}_{\text{B}}\text{T}$.}
\label{pearson}
\end{figure*}
\begin{figure*} 
\subfloat{\includegraphics[scale=0.5]{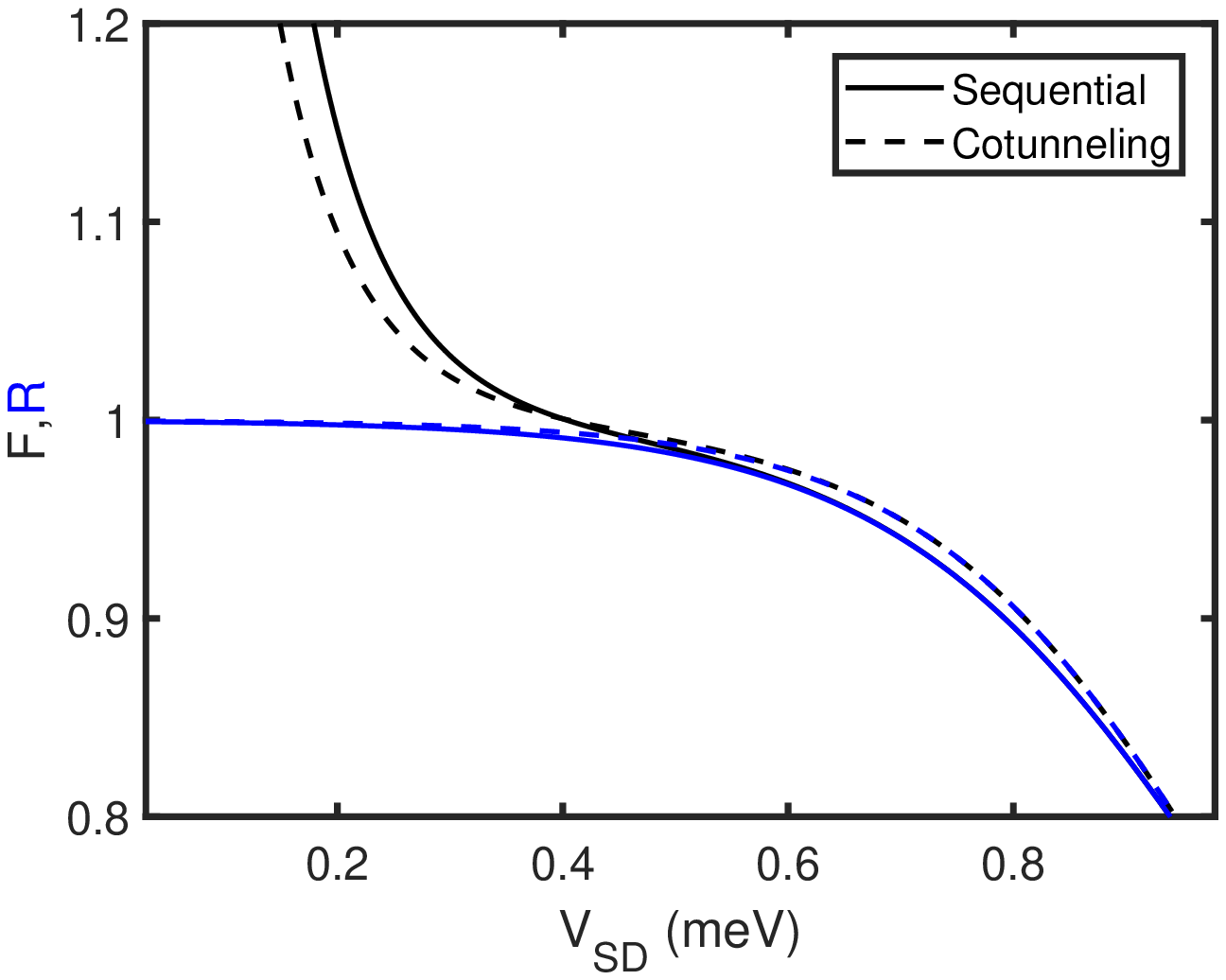}}
\subfloat{\includegraphics[scale=0.5]{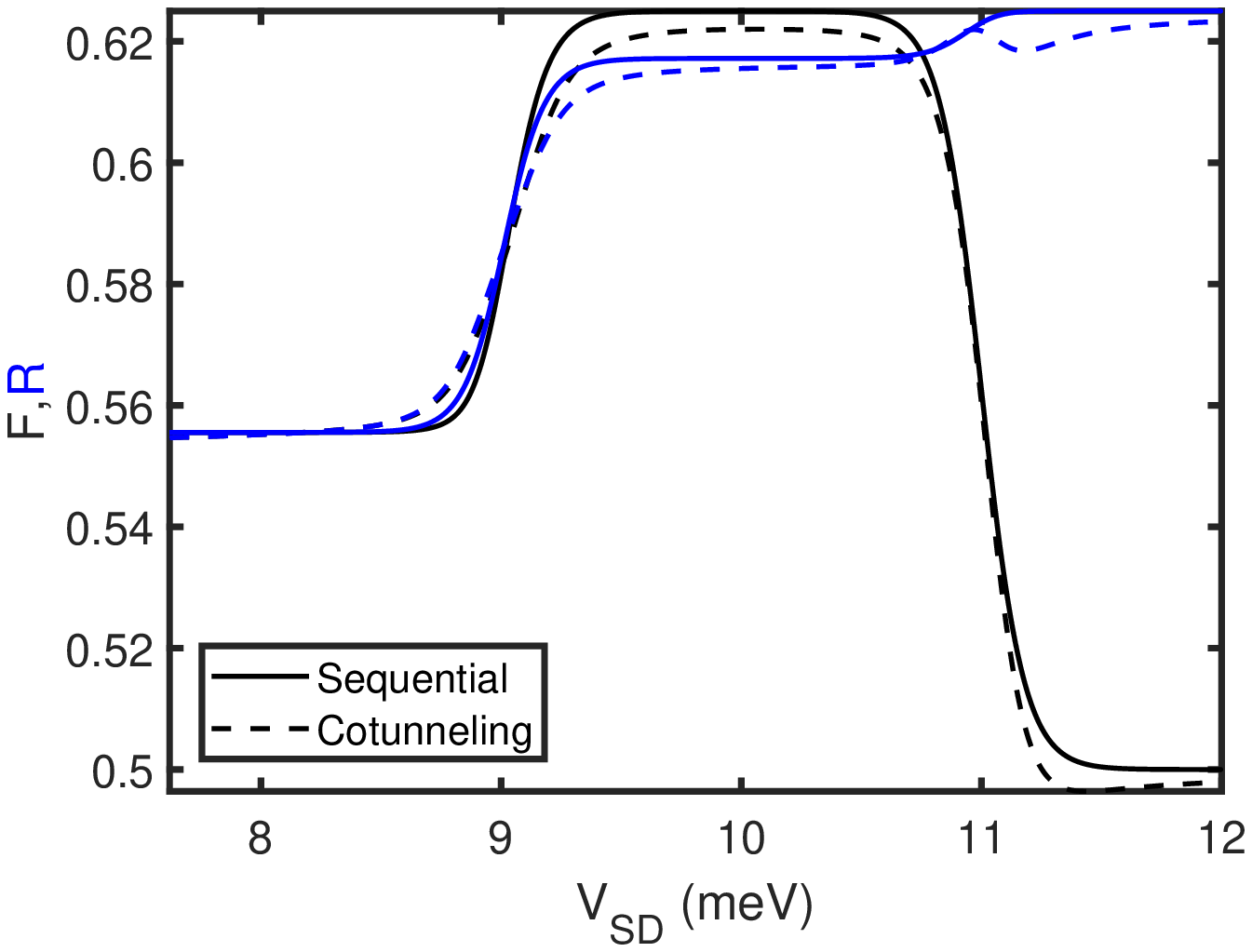}}
\caption{Colour online. Exact Fano factor $F$ and its prediction from waiting times under the renewal assumption $R$ over a range of voltages in (a) the Coulomb blockade regime and (b) the tunneling regime. The energies of the spin split electronic levels are $\varepsilon_{\uparrow}=0.5\text{meV}$ and $\varepsilon_{\downarrow}=1.5\text{meV}$, the Coulomb repulsion is $U=4\text{meV}$, $\text{k}_{\text{B}}\text{T}=75\mu\mbox{eV}$, and $\gamma=0.5\text{k}_{\text{B}}\text{T}$. }
\label{Relative_Dispersion}
\end{figure*}

Sequential tunneling through an Anderson impurity displays nonrenewal statistics in the high voltage regime, which is seen in Fig.(\ref{pearson}b). Here, due to the strong inelastic electron-electron interaction when a spin up and spin down electron are occupying the impurity, the correlation between subsequent waiting times is negative; a short waiting time is more likely to be followed by a long waiting time and vice versa. We note that the Coulomb repulsion is an order of magnitude greater than the electronic single-particle energies, so that if the system is doubly occupied it is likely for both electrons to subsequently tunnel out, which is a short waiting time, and then for the system to fill and empty again, which is a long waiting time. Thus, the nonrenewal behaviour does not arise from non-Markovian behaviour, as we work under the Markovian assumption, but rather from the multiple tunneling processes contained in the drain jump operator.\cite{Ptaszynski2017} Importantly, even though sequential processes dominate in this regime cotunneling still has an effect on the nonrenewal statisticis, slightly increasing the strength of the negative correlation between subsequent waiting times. 

Multiple authors have shown that when the renewal assumption is satisfied there is a direct link between the cumulants of the waiting time distribution and the current cumulants.\cite{Albert2011,Budini2011} Here, we focus on the Fano factor, which is the ratio of the zero-frequency noise to the average current:
\begin{align}
F & = \frac{S(0)}{2e\langle I \rangle} \\
& = \frac{\langle n^{2}\rangle - \langle n\rangle^{2}}{\langle n \rangle}.
\end{align}
The Fano factor in terms of waiting times is given by the randomness parameter:\cite{Ptaszynski2017,Albert2011,Budini2011}
\begin{align}
R & = \frac{\langle\tau^{2}\rangle-\langle\tau\rangle^{2}}{\langle\tau\rangle^{2}}.
\end{align}
If the renewal assumption holds, then $F=R$. Indeed, in Fig.(\ref{Relative_Dispersion}b) one can see that the two parameters diverge at the same voltage that the sequential correlation coefficient becomes nonzero, and that the difference between the $F$ and $R$ increases as the correlation increases. Furthermore, when cotunneling processes are included, $F$ and $R$ diverge at a later voltage, following the behaviour of the cotunneling correlation.

Since multiple cotunneling rates appear in the drain jump operator, one might expect that nonrenewal behaviour could be observed even in the Coulomb blockade regime when the strong Coulomb repulsion does not play a part in the transport. Fig.(\ref{pearson}a) shows that for small voltages the correlation is nonzero but as with a SRL the magnitude of the correlations are negligibly small. This is apparent in Fig.(\ref{Relative_Dispersion}a); the presence of cotunneling changes the Fano factor and randomness parameter from their sequential values, but they still are not visibly different. Note that the divergence of the Fano factor at zero voltage is due to the complete suppression of the Poissonian shot noise in comparison to the thermal noise.

So far we have shown plots that are either deep in the Coulomb blockade regime or well in the tunneling regime. This is because for certain voltage ranges between these two extremes the approach produces unphysically negative probability densities for small waiting times. For a SRL these unphysical WTDs clearly occur when $\varepsilon<\text{eV}$ and second order contributions actually reduce the total current, which amounts to $\text{\it{negative}}$ regularised cotunneling rates. From the point of view of the theory the total transition rate $\Gamma_{00}=\frac{\gamma}{2}n_{F}(\varepsilon-\mu_{S})(1-n_{F}(\varepsilon-\mu_{D})) + \Gamma^{(2)}$ is still positive, but $\Gamma^{(2)}$ can be negative.\cite{Koch2004,Koch2006} In such a regime and for small $\tau$ ($\sim10^{3}\text{fs}$) the WTD for a SRL is negative, which is shown in Eq.\eqref{short_time_SRL} when $\Gamma^{(2)}<0$.

For an Anderson impurity the situation is more complex; it appears that when at voltages where cotunneling processes dramatically decrease the sequential current the WTD is negative for small $\tau$. It is not yet clear how to resolve this interesting pathology; evidently there should be a well-defined WTD for all voltage ranges. This positivity violation could be an artefact of only going to second-order perturbation theory; on the other hand, it may not be physically correct to include negative rates at all in the definition of the jump operators. 

\section{Conclusion} \label{Conclusion}

In this paper we have extended the Markovian master equation technique for calculating WTDs in quantum electron transport to include cotunneling effects, and demonstrated the method for transport through a SRL and an Anderson impurity. Additionally, we have demonstrated that, similarly to the WTD for sequential tunneling through a SRL, the cotunneling WTD in Laplace space provides information on the individual source-drain couplings. Of particular interest is how cotunneling processes affect the nonrenwal statistics already present in the Anderson impurity, where electrons experience strong inelastic electron-electron interactions. We have shown that for large voltages cotunneling increases the magnitude of the non-negligible negative correlation between waiting times of subsequent electron tunnelings to the drain, which is caused by a strong electron-electron interaction, and thus increases the nonrenewal behaviour shown by the difference in the Fano factor and the randomness parameter. However, in the Coulomb blockade regime where cotunneling processes dominate, the correlation between subsequent waiting times is negligible and the system displays renewal behaviour.

\begin{acknowledgements}
We thank the two anonymous reviewers whose insightful comments and suggestions helped significantly improve this paper. This work was supported by an Australian Government Research Training Program Stipend Scholarship to SLR.
\end{acknowledgements}

\appendix 

\section{Cotunneling rates} \label{Cotunneling rates}

In this appendix we derive the cotunneling rates shown in Eq.\eqref{Cotunneling Elastic Final Rate}, Eq.\eqref{Cotunneling Inelastic Final Rate}, and Eq.\eqref{Cotunneling SRL final rate}, from the starting point of Eq.\eqref{Cotunneling general}. We note that the derivation generally follows the regularisation procedure detailed by Koch et al.\cite{Koch2004, Koch2006} As an example, consider the case of elastic tunneling through an initially empty dot. The cotunneling rate from the source to the drain is

\begin{widetext}
\begin{equation}
\Gamma_{00}^{SD} = 2\pi\lim_{\eta\rightarrow 0^{+}}\sum_{\sigma}\sum_{i,f}\big{|}\bra{f}\bra{0}H^{D,\sigma}_{T}\frac{1}{E_{i,0}-H_{0}+i\eta}H^{S,\sigma}_{T}\ket{0}\ket{i}\big{|}^{2} W_{i,0}^{S}W_{i,0}^{D}\times\delta(\varepsilon_{i,s}-\varepsilon_{f,d}),
\end{equation}
\end{widetext}
where we have summed over $\sigma$ to account for tunneling through either the $\uparrow$ or $\downarrow$ level. The initial state of the dot is $\ket{0}\otimes\ket{i}$ and the final state is $a_{\sigma}a^{\dagger}_{\sigma}\ket{0}\otimes a_{\nu_{D}}^{\dagger}a_{\nu_{S}}|i\rangle$. To span the possible configurations after the cotunneling the rate is summed over the electrode states $\nu_{S}$ and $\nu_{D}$. Additionally, it is assumed that the metal electrodes' density of states is constant. With these assumptions, the rate reduces to 

\begin{widetext}
\begin{equation} \label{unregularised rate integral version_specific_empty}
\Gamma_{00}^{SD} = \frac{\gamma^{S}\gamma^{D}}{2\pi}\lim_{\eta\rightarrow0^{+}}\int d\varepsilon \Big| \frac{1}{\varepsilon-\varepsilon_{\uparrow}+i\eta} + \frac{1}{\varepsilon-\varepsilon_{\downarrow}+i\eta}\Big| n_{F}(\varepsilon-\mu_{S})[1-n_{F}(\varepsilon-\mu_{D})].
\end{equation}
\end{widetext}

In general then, elastic cotunneling rates have the form 
\begin{widetext}
\begin{equation} \label{unregularised rate integral version_general}
\Gamma_{nn}^{SD} = \frac{\gamma^{S}\gamma^{D}}{2\pi}\lim_{\eta\rightarrow0^{+}}\int d\varepsilon \Big| \frac{1}{\varepsilon-E_{1}+i\eta} \pm \frac{1}{\varepsilon-E_{2}+i\eta}\Big| n_{F}(\varepsilon-\mu_{S})[1-n_{F}(\varepsilon-\mu_{D})],
\end{equation}
\end{widetext}
where $E_{1}\text{ and }E_{2}$ are derived from the cotunneling pathways involved in the rate, and the $\pm$ is only positive for elastic tunneling through an initially empty or initially doubly occupied system. Similarly, inelastic cotunneling rates have the general form 
\begin{widetext}
\begin{equation} \label{unregularised rate integral version_general_inelastic}
\Gamma_{\bar{\sigma}\sigma}^{\alpha\beta} = \frac{\gamma^{\alpha}\gamma^{\beta}}{2\pi}\lim_{\eta\rightarrow0^{+}}\int d\varepsilon \Big| \frac{1}{\varepsilon-\varepsilon_{\bar{\sigma}}-U+i\eta} - \frac{1}{\varepsilon-\varepsilon_{\bar{\sigma}}-i\eta}\Big| n_{F}(\varepsilon-\mu_{S})[1-n_{F}(\varepsilon-\mu_{D}+\varepsilon_{\sigma}-\varepsilon_{\bar{\sigma}})],
\end{equation}
\end{widetext}
The expanded form of either Eq.\eqref{unregularised rate integral version_general} or Eq.\eqref{unregularised rate integral version_general} consists of two square terms and the real component of the cross term: for example,

\begin{widetext}
\begin{align} \label{unregularised rate integral version_general_expanded}
\Gamma_{nn}^{SD} = & \frac{\gamma^{S}\gamma^{D}}{2\pi}\lim_{\eta\rightarrow0^{+}}\Bigg[\int d\varepsilon \frac{1}{(\varepsilon-E_{1})^{2}+\eta^{2}} n_{F}(\varepsilon-\mu_{S})[1-n_{F}(\varepsilon-\mu_{D})] \nonumber \\
& + \int d\varepsilon \frac{1}{(\varepsilon-E_{2})^{2}+\eta^{2}} n_{F}(\varepsilon-\mu_{S})[1-n_{F}(\varepsilon-\mu_{D})] \nonumber \\
& \pm 2\Re\int d\varepsilon \frac{1}{\varepsilon-E_{1}+i\eta} \cdot \frac{1}{\varepsilon-E_{2}-i\eta} n_{F}(\varepsilon-\mu_{S})[1-n_{F}(\varepsilon-\mu_{D})\Bigg]. \nonumber \\
\end{align}
\end{widetext}

It has been noted in the literature that the intermediate virtual state in the dot has a finite width, which is proportional to the coupling strength $\eta\sim\gamma$, and so a divergence is avoided in the denominator of the integrand.\cite{Averin1994,Turek2002,Koch2006,Koch2004} Additionally, the two square terms in the overall rate include not only the rate of that particular event from cotunneling, but also the contribution to that process from sequential tunneling, as a cotunneling event can be mimicked by two sequential tunneling events. Thus, it is necessary to remove this sequential overcounting by expanding the integrands of the first two integrals in Eq.\eqref{unregularised rate integral version_general_expanded} in a power series about $\eta$, and discarding the $\eta^{-1}$ term as with the $\gamma^{S}\gamma^{D}$ prefactor it is overall $\mathcal{O}(\gamma)$. For a simple system such as a single resonant level some groups choose to remove the overcounting and then compute the rate numerically with Cauchy's principal value.\cite{Turek2002} However, we follow the approach of Koch et al. and evaluate the integral analytically by transforming it to a contour integral over a semicircle in the upper half-plane of complex space and using residue theory. The final expressions for the elastic and inelastic rates are given in Eq.\eqref{Cotunneling Elastic Final Rate} and Eq.\eqref{Cotunneling Inelastic Final Rate}, where the trigamma functions $\psi^{(1)}(x)$ come from the squared terms in the rate, the digamma functions $\psi(x)$ come from the cross term, and both originate from the complex poles of the Fermi-Dirac distributions $n_{F}(\varepsilon-\mu_{S/D})$, known as the Matsubara frequencies.

The cotunneling rate for a SRL is a simpler process, as there is only one pathway through the system, and so only the squared term appears in the expanded integral.

\bibliographystyle{unsrt}
\bibliography{Main_text_incl._figures}

\begin{thebibliography}{10}

\bibitem{Tan2017}
Kuan~Yen Tan, Matti Partanen, Russell~E. Lake, Joonas Govenius, Shumpei Masuda,
  and Mikko M{\"{o}}tt{\"{o}}nen.
\newblock {Quantum-circuit refrigerator}.
\newblock {\em Nature Communications}, 8:15189, 2017.

\bibitem{Bogani2008}
Lapo Bogani and Wolfgang Wernsdorfer.
\newblock {Molecular spintronics using single-molecule magnets}.
\newblock {\em Nature Materials}, 7(3):179--186, 2008.

\bibitem{Xiang2016}
Dong Xiang, Xiaolong Wang, Chuancheng Jia, Takhee Lee, and Xuefeng Guo.
\newblock {Molecular-Scale Electronics: From Concept to Function}.
\newblock {\em Chemical Reviews}, 116(7):4318--4440, 2016.

\bibitem{Averin1989}
D.~V. Averin and A.~A. Odintsov.
\newblock {Macroscopic quantum tunneling of the electric charge in small tunnel
  junctions}.
\newblock {\em Physics Letters A}, 140(5):251--257, 1989.

\bibitem{PhysRevLett.65.3037}
L~J Geerligs, D~V Averin, and J~E Mooij.
\newblock {Observation of macroscopic quantum tunneling through the Coulomb
  energy barrier}.
\newblock {\em Phys. Rev. Lett.}, 65(24):3037--3040, 1990.

\bibitem{Averin1990}
D.~V. Averin and Yu~V. Nazarov.
\newblock {Virtual electron diffusion during quantum tunneling of the electric
  charge}.
\newblock {\em Phys. Rev. Lett.}, 65(19):2446--2449, 1990.

\bibitem{scheer2010molecular}
J.~Cuevas and E.~Scheer.
\newblock {\em {Molecular Electronics: An Introduction to Theory and
  Experiment}}.
\newblock World Scientific series in nanoscience and nanotechnology. World
  Scientific Publishing Company Pte Limited, 2010.

\bibitem{Bruus2002}
Henrik Bruus and Karsten Flensberg.
\newblock {\em {Many-body quantum theory in condensed matter physics: An
  Introduction}}.
\newblock 2002.

\bibitem{Beenakker1991}
C.~W~J Beenakker.
\newblock {Theory of Coulomb-blockade oscillations in the conductance of a
  quantum dot}.
\newblock {\em Phys. Rev. B}, 44(4):1646--1656, 1991.

\bibitem{Romito2014}
Alessandro Romito and Yuval Gefen.
\newblock {Weak measurement of cotunneling time}.
\newblock {\em Phys. Rev. B}, 90(8):085417, 2014.

\bibitem{PhysRevLett.86.878}
S~{De Franceschi}, S~Sasaki, J~M Elzerman, W~G van~der Wiel, S~Tarucha, and L~P
  Kouwenhoven.
\newblock {Electron Cotunneling in a Semiconductor Quantum Dot}.
\newblock {\em Phys. Rev. Lett.}, 86(5):878--881, 2001.

\bibitem{Koch2006}
Jens Koch, Felix {Von Oppen}, and A.~V. Andreev.
\newblock {Theory of the Franck-Condon blockade regime}.
\newblock {\em Phys. Rev. B}, 74(20):205438, 2006.

\bibitem{Koch2004}
Jens Koch, Felix {Von Oppen}, Yuval Oreg, and Eran Sela.
\newblock {Thermopower of single-molecule devices}.
\newblock {\em Phys. Rev. B}, 70(19):195107, 2004.

\bibitem{Turek2002}
M.~Turek and K.~A. Matveev.
\newblock {Cotunneling thermopower of single electron transistors}.
\newblock {\em Phys. Rev. B}, 65(11):115332, 2002.

\bibitem{Dinaii2014}
Yehuda Dinaii, Alexander Shnirman, and Yuval Gefen.
\newblock {Statistics of energy dissipation in a quantum dot operating in the
  cotunneling regime}.
\newblock {\em Phys. Rev. B}, 90(20):201404, 2014.

\bibitem{Gergs2015}
Niklas~M. Gergs, Christoph B.~M. H{\"{o}}rig, Maarten~R. Wegewijs, and Dirk
  Schuricht.
\newblock {Charge fluctuations in nonlinear heat transport}.
\newblock {\em Phys. Rev. B}, 91(20):201107, 2015.

\bibitem{Golovach2004}
Vitaly~N. Golovach and Daniel Loss.
\newblock {Transport through a double quantum dot in the sequential tunneling
  and cotunneling regimes}.
\newblock {\em Phys. Rev. B}, 69(24):245327, 2004.

\bibitem{Pedersen2007}
Jonas~Nyvold Pedersen, Benny Lassen, Andreas Wacker, and Matthias~H. Hettler.
\newblock {Coherent transport through an interacting double quantum dot: Beyond
  sequential tunneling}.
\newblock {\em Phys. Rev. B}, 75(23):235314, 2007.

\bibitem{Begemann2010}
Georg Begemann, Sonja Koller, Milena Grifoni, and Jens Paaske.
\newblock {Inelastic cotunneling in quantum dots and molecules with weakly
  broken degeneracies}.
\newblock {\em Phys. Rev. B}, 82(4):045316, 2010.

\bibitem{Leijnse2009}
M.~Leijnse, M.~R. Wegewijs, and M.~H. Hettler.
\newblock {Pair tunneling resonance in the single-electron transport regime}.
\newblock {\em Phys. Rev. Lett.}, 103(15):156803, 2009.

\bibitem{Leijnse2008}
M.~Leijnse and M.~R. Wegewijs.
\newblock {Kinetic equations for transport through single-molecule
  transistors}.
\newblock {\em Phys. Rev. B}, 78(23):235424, 2008.

\bibitem{PhysRevB.91.235413}
Kristen Kaasbjerg and Wolfgang Belzig.
\newblock {Full counting statistics and shot noise of cotunneling in quantum
  dots and single-molecule transistors}.
\newblock {\em Phys. Rev. B}, 91(23):235413, 2015.

\bibitem{PhysRevB.80.235306}
Clive Emary.
\newblock {Counting statistics of cotunneling electrons}.
\newblock {\em Phys. Rev. B}, 80(23):235306, 2009.

\bibitem{PhysRevLett.95.146806}
Axel Thielmann, Matthias~H Hettler, J{\"{u}}rgen K{\"{o}}nig, and Gerd
  Sch{\"{o}}n.
\newblock {Cotunneling Current and Shot Noise in Quantum Dots}.
\newblock {\em Phys. Rev. Lett.}, 95(14):146806, 2005.

\bibitem{Sukhorukov2000}
Eugene~V. Sukhorukov, Guido Burkard, Daniel Loss, and D.~K. Young.
\newblock {Noise of a quantum dot system in the cotunneling regime}.
\newblock {\em Phys. Rev. B}, 63(12):125315, 2001.

\bibitem{Thielmann2005}
Axel Thielmann, Matthias~H. Hettler, J{\"{u}}rgen K{\"{o}}nig, and Gerd
  Sch{\"{o}}n.
\newblock {Super-Poissonian noise, negative differential conductance, and
  relaxation effects in transport through molecules, quantum dots, and
  nanotubes}.
\newblock {\em Phys. Rev. B}, 71(4):045341, 2005.

\bibitem{Weymann2008}
I.~Weymann, J.~Barna{\'{s}}, and S.~Krompiewski.
\newblock {Transport through single-wall metallic carbon nanotubes in the
  cotunneling regime}.
\newblock {\em Phys. Rev. B}, 78(3):035422, 2008.

\bibitem{Carmi2012}
Assaf Carmi and Yuval Oreg.
\newblock {Enhanced shot noise in asymmetric interacting two-level systems}.
\newblock {\em Phys. Rev. B}, 85(4):045325, 2012.

\bibitem{Aghassi2008}
Jasmin Aghassi, Matthias~H. Hettler, and Gerd Sch{\"{o}}n.
\newblock {Cotunneling assisted sequential tunneling in multilevel quantum
  dots}.
\newblock {\em Applied Physics Letters}, 92(20):202101, 2008.

\bibitem{Braggio2006}
Alessandro Braggio, J{\"{u}}rgen K{\"{o}}nig, and Rosario Fazio.
\newblock {Full counting statistics in strongly interacting systems:
  Non-Markovian effects}.
\newblock {\em Phys. Rev. Lett.}, 96(2):026805, 2006.

\bibitem{Utsumi2006}
Yasuhiro Utsumi, Dmitri~S. Golubev, and Gerd Sch{\"{o}}n.
\newblock {Full counting statistics for a single-electron transistor:
  Nonequilibrium effects at intermediate conductance}.
\newblock {\em Phys. Rev. Lett.}, 96(8):086803, 2006.

\bibitem{OKazaki2013}
Yuma Okazaki, Satoshi Sasaki, and Koji Muraki.
\newblock {Shot noise spectroscopy on a semiconductor quantum dot in the
  elastic and inelastic cotunneling regimes}.
\newblock {\em Phys. Rev. B}, 87(4):041302, 2013.

\bibitem{Zhang2007}
Yiming Zhang, L.~DiCarlo, D.~T. McClure, M.~Yamamoto, S.~Tarucha, C.~M. Marcus,
  M.~P. Hanson, and A.~C. Gossard.
\newblock {Noise correlations in a Coulomb-blockaded quantum dot}.
\newblock {\em Phys. Rev. Lett.}, 99(3):036603, 2007.

\bibitem{ONac2005}
E.~Onac, F.~Balestro, B.~Trauzettel, C.~F~J Lodewijk, and L.~P. Kouwenhoven.
\newblock {Shot-noise detection in a carbon nanotube quantum dot}.
\newblock {\em Phys. Rev. Lett.}, 96(2):026803, 2006.

\bibitem{Brandes2008}
Tobias Brandes.
\newblock {Waiting Times and Noise in Single Particle Transport}.
\newblock {\em Ann. Phys. (Berlin)}, 17(7):477--496, 2008.

\bibitem{PhysRevA.39.1200}
H~J Carmichael, Surendra Singh, Reeta Vyas, and P~R Rice.
\newblock {Photoelectron waiting times and atomic state reduction in resonance
  fluorescence}.
\newblock {\em Phys. Rev. A}, 39(3):1200--1218, 1989.

\bibitem{Srinivas2010}
M.D. Srinivas and E.B. Davies.
\newblock {Photon Counting Probabilities in Quantum Optics}.
\newblock {\em Optica Acta: International Journal of Optics}, 28(7):981--996,
  2010.

\bibitem{Thomas2013}
Konrad~H. Thomas and Christian Flindt.
\newblock {Electron waiting times in non-Markovian quantum transport}.
\newblock {\em Phys. Rev. B}, 87(12):121405, 2013.

\bibitem{Rudge2016}
Samuel~L. Rudge and Daniel~S. Kosov.
\newblock {Distribution of residence times as a marker to distinguish different
  pathways for quantum transport}.
\newblock {\em Phys. Rev. E}, 94(4):042134, 2016.

\bibitem{Rudge2016a}
Samuel~L. Rudge and Daniel~S. Kosov.
\newblock {Distribution of tunnelling times for quantum electron transport}.
\newblock {\em Journal of Chemical Physics}, 144(12):124105, 2016.

\bibitem{Kosov2016}
Daniel~S. Kosov.
\newblock {Waiting time distribution for electron transport in a molecular
  junction with electron-vibration interaction}.
\newblock {\em Journal of Chemical Physics}, 146(7):074102, 2017.

\bibitem{Dasenbrook2015}
David Dasenbrook, Patrick~P. Hofer, and Christian Flindt.
\newblock {Electron waiting times in coherent conductors are correlated}.
\newblock {\em Phys. Rev. B}, 91(19):195420, 2015.

\bibitem{Albert2011}
Mathias Albert, Christian Flindt, and Markus B{\"{u}}ttiker.
\newblock {Distributions of waiting times of dynamic single-electron emitters}.
\newblock {\em Phys. Rev. Lett.}, 107(8):086805, 2011.

\bibitem{Albert2012}
Mathias Albert, G\'eraldine Haack, Christian Flindt, and Markus B\"uttiker.
\newblock Electron waiting times in mesoscopic conductors.
\newblock {\em Phys. Rev. Lett.}, 108(18):186806, May 2012.

\bibitem{Ubbelohde2012}
Niels Ubbelohde, Christian Fricke, Christian Flindt, Frank Hohls, and Rolf~J.
  Haug.
\newblock {Measurement of finite-frequency current statistics in a
  single-electron transistor}.
\newblock {\em Nature Communications}, 3:612, 2012.

\bibitem{Lu2003}
Wei Lu, Zhongqing Ji, Loren Pfeiffer, K~W West, and A~J Rimberg.
\newblock {Real-time detection of electron tunnelling in a quantum dot}.
\newblock {\em Nature}, 423(6938):422--425, 2003.

\bibitem{Gustavsson2009}
S.~Gustavsson, R.~Leturcq, M.~Studer, I.~Shorubalko, T.~Ihn, K.~Ensslin, D.~C.
  Driscoll, and A.~C. Gossard.
\newblock {Electron counting in quantum dots}.
\newblock {\em Surface Science Reports}, 64(6):191--232, 2009.

\bibitem{Gustavsson2008a}
S.~Gustavsson, R.~Leturcq, M.~Studer, T.~Ihn, K.~Ensslin, D.~C. Driscoll, and
  A.~C. Gossard.
\newblock {Time-resolved detection of single-electron interference}.
\newblock {\em Nano Letters}, 8(8):2547--2550, 2008.

\bibitem{PhysRevB.78.155309}
S~Gustavsson, M~Studer, R~Leturcq, T~Ihn, K~Ensslin, D~C Driscoll, and A~C
  Gossard.
\newblock {Detecting single-electron tunneling involving virtual processes in
  real time}.
\newblock {\em Phys. Rev. B}, 78(15):155309, 2008.

\bibitem{Zilberberg2014}
Oded Zilberberg, Assaf Carmi, and Alessandro Romito.
\newblock {Measuring cotunneling in its wake}.
\newblock {\em Phys. Rev. B}, 90(20):205413, 2014.

\bibitem{Haack2015}
G~Haack, A~Steffens, J~Eisert, and R~Hübener.
\newblock Continuous matrix product state tomography of quantum transport
  experiments.
\newblock {\em New Journal of Physics}, 17(11):113024, 2015.

\bibitem{VanKampen1981}
N~G {Van Kampen}.
\newblock {\em {Stochastic processes in physics and chemistry}}, volume~11.
\newblock 1992.

\bibitem{Rajabi2013}
Leila Rajabi, Christina P{\"{o}}ltl, and Michele Governale.
\newblock {Waiting time distributions for the transport through a quantum-dot
  tunnel coupled to one normal and one superconducting lead}.
\newblock {\em Phys. Rev. Lett.}, 111(6):067002, 2013.

\bibitem{Chevallier2016}
D.~Chevallier, M.~Albert, and P.~Devillard.
\newblock {Probing Majorana and Andreev bound states with waiting times}.
\newblock {\em Europhysics Letters}, 116(2):27005, 2016.

\bibitem{Walldorf2018}
Nicklas Walldorf, Ciprian Padurariu, Antti~Pekka Jauho, and Christian Flindt.
\newblock {Electron Waiting Times of a Cooper Pair Splitter}.
\newblock {\em Phys. Rev. Lett.}, 120(8):087701, 2018.

\bibitem{Sothmann2014}
Bj{\"{o}}rn Sothmann.
\newblock {Electronic waiting-time distribution of a quantum-dot spin valve}.
\newblock {\em Phys. Rev. B}, 90(15):155315, 2014.

\bibitem{Welack2009}
S.~Welack, S.~Mukamel, and Yi~Jing Yan.
\newblock {Waiting time distributions of electron transfers through quantum dot
  Aharonov-Bohm interferometers}.
\newblock {\em Europhysics Letters}, 85(5):57008, 2009.

\bibitem{Ptaszynski2017}
Krzysztof Ptaszy{\'{n}}ski.
\newblock {Nonrenewal statistics in transport through quantum dots}.
\newblock {\em Phys. Rev. B}, 95(4):045306, 2017.

\bibitem{Haack2014}
G\'eraldine Haack, Mathias Albert, and Christian Flindt.
\newblock Distributions of electron waiting times in quantum-coherent
  conductors.
\newblock {\em Phys. Rev. B}, 90(20):205429, 2014.

\bibitem{Albert2016}
M.~Albert, D.~Chevallier, and P.~Devillard.
\newblock Waiting times of entangled electrons in normal–superconducting
  junctions.
\newblock {\em Physica E}, 76:209 -- 215, 2016.

\bibitem{Albert2014}
M.~Albert and P.~Devillard.
\newblock {Waiting time distribution for trains of quantized electron pulses}.
\newblock {\em Phys. Rev. B}, 90(3):035431, 2014.

\bibitem{Tang2014}
Gao~Min Tang, Fuming Xu, and Jian Wang.
\newblock {Waiting time distribution of quantum electronic transport in the
  transient regime}.
\newblock {\em Phys. Rev. B}, 89(20):205310, 2014.

\bibitem{Tang2014a}
Gao~Min Tang and Jian Wang.
\newblock {Full-counting statistics of charge and spin transport in the
  transient regime: A nonequilibrium Green's function approach}.
\newblock {\em Phys. Rev. B}, 90(19):195422, 2014.

\bibitem{PhysRevB.92.125435}
R~{Seoane Souto}, R~Avriller, R~C Monreal,
  A~Mart$\backslash$'$\backslash$in-Rodero, and A~{Levy Yeyati}.
\newblock {Transient dynamics and waiting time distribution of molecular
  junctions in the polaronic regime}.
\newblock {\em Phys. Rev. B}, 92(12):125435, 2015.

\bibitem{Gurvitz1996}
S.~A. Gurvitz, H.~J. Lipkin, and Ya~S. Prager.
\newblock {Interference effects in resonant tunneling and the Pauli principle}.
\newblock {\em Physics Letters, Section A: General, Atomic and Solid State
  Physics}, 212(1-2):91--96, 1996.

\bibitem{PhysRevB.71.205304}
Xin-Qi Li, Junyan Luo, Yong-Gang Yang, Ping Cui, and Yi-Jing Yan.
\newblock Quantum master-equation approach to quantum transport through
  mesoscopic systems.
\newblock {\em Phys. Rev. B}, 71(20):205304, May 2005.

\bibitem{PhysRevB.74.235309}
Upendra Harbola, Massimiliano Esposito, and Shaul Mukamel.
\newblock Quantum master equation for electron transport through quantum dots
  and single molecules.
\newblock {\em Phys. Rev. B}, 74(23):235309, Dec 2006.

\bibitem{dzhioev11a}
A.~A. Dzhioev and D.~S. Kosov.
\newblock Super-fermion representation of quantum kinetic equations for the
  electron transport problem.
\newblock {\em J. Chem. Phys.}, 134:044121, 2011.

\bibitem{kosov18}
Daniel~S. Kosov.
\newblock Telegraph noise in markovian master equation for electron transport
  through molecular junctions.
\newblock {\em The Journal of Chemical Physics}, 148(18):184108, 2018.

\bibitem{Nazarov1993}
Yuli~V. Nazarov.
\newblock {Quantum interference, tunnel junctions and resonant tunneling
  interferometer}.
\newblock {\em Physica B: Physics of Condensed Matter}, 189(1-4):57--69, 1993.

\bibitem{Tu2008}
Matisse~W.Y. Tu and Wei~Min Zhang.
\newblock {Non-Markovian decoherence theory for a double-dot charge qubit}.
\newblock {\em Phys. Rev. B}, 78(23):235311, 2008.

\bibitem{kosov13}
Daniel~S Kosov, Toma{\v z} Prosen, and Bojan {\v Z}unkovi{\v c}.
\newblock A markovian kinetic equation approach to electron transport through a
  quantum dot coupled to superconducting leads.
\newblock {\em Journal of Physics: Condensed Matter}, 25(7):075702, 2013.

\bibitem{PhysRevB.87.155439}
Sebastian Pfaller, Andrea Donarini, and Milena Grifoni.
\newblock Subgap features due to quasiparticle tunneling in quantum dots
  coupled to superconducting leads.
\newblock {\em Phys. Rev. B}, 87:155439, Apr 2013.

\bibitem{Kohler2005}
Sigmund Kohler, J{\"{o}}rg Lehmann, and Peter H{\"{a}}nggi.
\newblock {Driven quantum transport on the nanoscale}.
\newblock {\em Physics Reports}, 406(6):379--443, 2005.

\bibitem{Gurvitz1998}
S.~Gurvitz.
\newblock {Rate equations for quantum transport in multidot systems}.
\newblock {\em Phys. Rev. B}, 57(11):6602--6611, 1998.

\bibitem{Gurvitz1996a}
S.~Gurvitz and Ya~S. Prager.
\newblock {Microscopic derivation of rate equations for quantum transport}.
\newblock {\em Phys. Rev. B}, 53(23):15932--15943, 1996.

\bibitem{Davies1993}
John~H. Davies, Selman Hershfield, Per Hyldgaard, and John~W. Wilkins.
\newblock {Current and rate equation for resonant tunneling}.
\newblock {\em Phys. Rev. B}, 47(8):4603--4618, 1993.

\bibitem{Li2005}
Xin-Qi Li, Junyan Luo, Yong-Gang Yang, Ping Cui, and Yi-Jing Yan.
\newblock {Quantum master-equation approach to quantum transport through
  mesoscopic systems}.
\newblock {\em Phys. Rev. B}, 71(20):205304, 2005.

\bibitem{Emary2007}
C.~Emary, D.~Marcos, R.~Aguado, and T.~Brandes.
\newblock {Frequency-dependent counting statistics in interacting nanoscale
  conductors}.
\newblock {\em Phys. Rev. B}, 76:161404, 2007.

\bibitem{Averin1994}
D.~V. Averin.
\newblock {Periodic conductance oscillations in the single-electron tunneling
  transistor}.
\newblock {\em Physica B: Physics of Condensed Matter}, 194-196(PART
  1):979--980, 1994.

\bibitem{Schoeller1994}
Herbert Schoeller and Gerd Sch{\"{o}}n.
\newblock {Mesoscopic quantum transport: Resonant tunneling in the presence of
  a strong Coulomb interaction}.
\newblock {\em Phys. Rev. B}, 50(24):18436--18452, 1994.

\bibitem{Konig1997}
J{\"{u}}rgen K{\"{o}}nig, Herbert Schoeller, and Gerd Sch{\"{o}}n.
\newblock {Cotunneling at resonance for the single-electron transistor}.
\newblock {\em Phys. Rev. Lett.}, 78(23):4482--4485, 1997.

\bibitem{PhysRevB.77.195416}
Carsten Timm.
\newblock {Tunneling through molecules and quantum dots: Master-equation
  approaches}.
\newblock {\em Phys. Rev. B}, 77(19):195416, 2008.

\bibitem{Nazarov1999}
Yuli~V Nazarov.
\newblock {Universalities of weak localization}.
\newblock {\em arXiv.org}, cond-mat.m:8143, 1999.

\bibitem{Bagrets2003}
D.~A. Bagrets and Yu~V. Nazarov.
\newblock {Full counting statistics of charge transfer in Coulomb blockade
  systems}.
\newblock {\em Phys. Rev. B}, 67(8):085316, 2003.

\bibitem{nonrenewal-kosov}
Daniel~S. Kosov.
\newblock Non-renewal statistics for electron transport in a molecular junction
  with electron-vibration interaction.
\newblock {\em The Journal of Chemical Physics}, 147(10):104109, 2017.

\bibitem{Budini2011}
Adri\'an~A. Budini.
\newblock Large deviations of ergodic counting processes: A statistical
  mechanics approach.
\newblock {\em Phys. Rev. E}, 84:011141, Jul 2011.

\end{thebibliography}

\end{document}